\documentclass[]{aa} 

\usepackage{multirow}
\usepackage{caption}
\usepackage{adjustbox}
\usepackage{subcaption}
\usepackage{graphicx}
\usepackage[varg]{txfonts}
\usepackage{natbib}
\usepackage{hyperref}
\usepackage[utf8]{inputenc}
\usepackage{xcolor}
\usepackage{textcomp}
\usepackage{lscape}
\usepackage{stfloats}
\usepackage{cleveref}
\usepackage[table,xcdraw]{xcolor}
\usepackage[T1]{fontenc}     
\usepackage{placeins}
\usepackage{diagbox}
\usepackage{babel}           
\usepackage{graphicx}
\usepackage{float} 
\usepackage[normalem]{ulem}

\newcommand{\vnh}{\hat{\mathbf n}}
\usepackage{hyperref}
\hypersetup{
  colorlinks,
  filecolor=blue,
  citecolor=blue,
  linkcolor=blue,
  urlcolor=blue}

\begin{document}

\title{Towards precision cosmology with Void $\times$ CMB correlations (II)} \subtitle{Impact of mock catalogs on the Void $\times$ CMB lensing signal.}

\titlerunning{Towards precision cosmology with Void x CMB correlations}
\author{%
Mar Pérez Sar\inst{1,2}\thanks{\href{mailto:perezsarmar@gmail.com}{perezsarmar@gmail.com} }
\and Carlos Hernández Monteagudo\inst{1,2}
\and András Kovács\inst{3,4}
\and Alice Pisani\inst{5,6}
\and Yun Wang\inst{7}}

\institute{%
Instituto de Astrofísica de Canarias, Calle Vía Láctea s/n, E-38205, La Laguna, Tenerife, Spain
\and
Departamento de Astrofísica, Universidad de La Laguna, E-38206, La Laguna, Tenerife, Spain
\and
MTA–CSFK Lendület “Momentum” Large-Scale Structure (LSS) Research Group, Konkoly Thege Miklós út 15-17, H-1121 Budapest, Hungary
\and
Konkoly Observatory, HUN-REN Research Centre for Astronomy and Earth Sciences, Budapest, Hungary
\and
CPPM, Aix-Marseille Université, CNRS/IN2P3, Marseille, France
\and
Department of Astrophysical Sciences, Peyton Hall, Princeton University, Princeton, NJ 08544, USA
\and 
IPAC, California Institute of Technology, Mail Code 314-6, 1200 E. California Blvd., Pasadena, CA 91125
}

   \date{Received xxx, 2024; accepted xxx, 2024}


\abstract
{Gravitational lensing by large-scale structure imprints secondary anisotropies on the Cosmic Microwave Background (CMB) that can be exploited to probe cosmology.  In particular, cosmic voids produce a characteristic lensing signature detectable through Void $\times$ CMB cross-correlations. This signal has been robustly measured in the past but its cosmological constraining power remains limited by the incomplete knowledge of how methodological choices affect its measurement and by its uncertain dependence on cosmological parameters. Using a set of validated \emph{Roman} mock catalogs, we first quantify how mock construction impacts the measured signal and then forecast the capabilities of \emph{Roman}, in combination with current and upcoming CMB surveys such as \emph{Planck}, SO and CMB-S4-like experiments. We analyze the signal-to-noise ratio (S/N) for different void definitions (2D and 3D), stacking approaches (rescaled versus non-rescaled profiles), CMB map filtering schemes and noise levels.
In contrast to galaxy and void statistics, we find that the Void $\times$ CMB lensing signal is less sensitive to the choice of mock catalog, indicating that future tensions with data are unlikely to stem from mock inaccuracies alone.  The highest S/N is achieved for 2D voids with rescaled profiles. We forecast S/N $\approx$ 13$\sigma$ (8$\sigma$) for 2D (3D) \emph{Roman} voids combined with \emph{Planck}, increasing to 22$\sigma$ (13$\sigma$) for SO and 31$\sigma$ (18$\sigma$) for CMB-S4-like surveys. For comparison, current \emph{Planck}-based measurements reach S/N $\approx$ 17$\sigma$ for 3D DESI Legacy Survey LRG voids and 5.9$\sigma$ for 2D DES Year 3 voids, both covering substantially larger areas than \emph{Roman} ($\sim$2000 deg$^2$).
While the cosmological dependence of this observable remains to be quantified, \emph{Roman} together with next-generation of LSS and CMB surveys opens a path toward the first direct cosmological constraints from Void $\times$ CMB lensing.
}



   \maketitle \nolinenumbers
%
\section{Introduction}
\label{sec:intro}

\begin{figure*}[h]
    \centering
    \includegraphics[width=19cm]{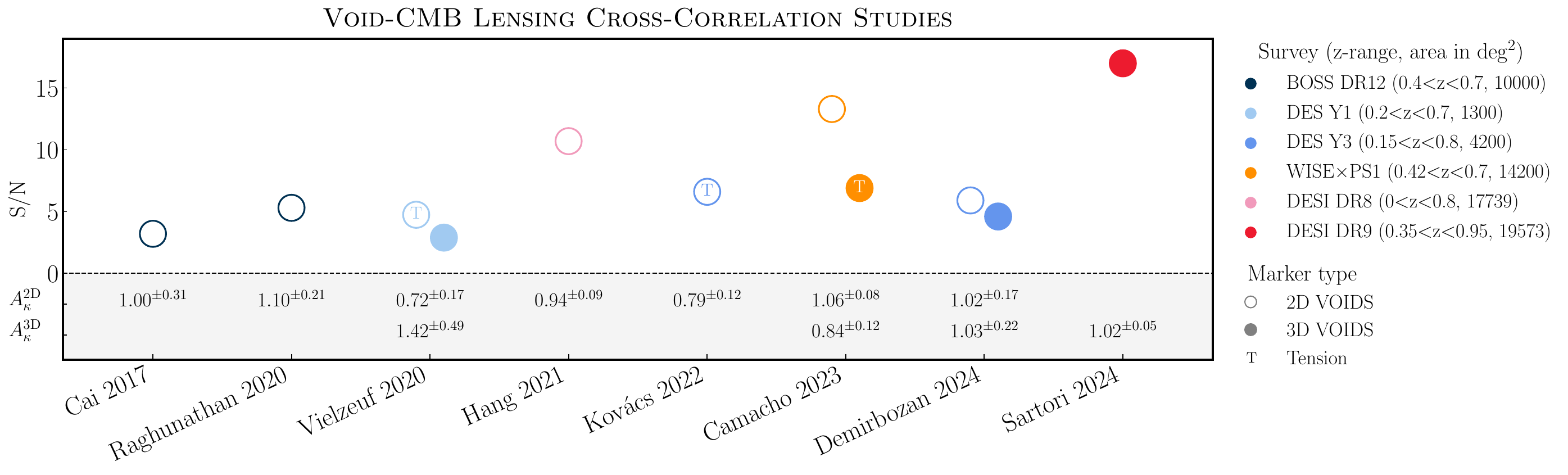} 
    \caption{Review of prior work on Void $\times$ CMB lensing cross-correlations. The figure presents the signal-to-noise ratio (S/N) achieved and the value of the parameter $\rm A_\kappa$
  which quantifies the agreement between the observed signal and cosmological simulations. Unfilled circles denote results from 2D voids and filled circles from 3D voids. A `T’ symbol inside a point indicates a tension with theoretical expectations for some configuration.}
    \label{fig:SoA_CMBxVoids}
\end{figure*}

In 1978, the same year the Nobel Committee for Physics recognized the Cosmic Microwave Background (CMB) as the first compelling evidence for the Big Bang model \citep{gamow1946,gamow1948,alpher&bethe&gamow1948,alpher&herman1948,penzias1965}, two independent studies \citep{joever&einasto1978,gregory&thompson1978} reported the first observational hints of cosmic voids, starting to uncover the "cellular pattern" now identified as the cosmic web within the large-scale structure (LSS). Nearly fifty years later, the CMB and LSS have become the two foundational pillars of modern cosmology and constitute unique probes to test the dynamics and contents of our Universe through its cosmic history \citep{dodelson2003,peebles2020}.

The cosmological information within the CMB is primarily encoded in its angular anisotropies---tiny fluctuations at the level of $\sim 10^{-5}~\mathrm{K}$ superimposed on the uniform background of the most perfect blackbody ever measured ($T_{\mathrm{CMB}} = 2.72548 \pm 0.00057 ~\mathrm{K}$, ~\citealt{COBE1992_survey,COBE1993_survey}). The anisotropies represent the seeds that would later grow into the cosmic web and are classified into two types: primary and secondary anisotropies. Primary anisotropies originate at the surface of last scattering ($z \sim 1100$) and reflect the physical processes at play in the early Universe, mainly the oscillation of the photon–baryon plasma in the evolving dark matter potential wells and the Thomson scattering keeping both components bound until recombination. Secondary anisotropies arise from interactions of the resultant CMB photons with the intervening large-scale structures through gravitational and scattering processes. Gravitational effects includes lensing, i.e, the deflection of CMB photons due to spatially varying gravitational potentials \citep{lewisandchallinor2006} and the Integrated Sachs–Wolfe (ISW) effect, i.e, the change of CMB photon energies due to time-varying gravitational potentials \citep{sachs&wolfe1967}. Scattering effects originate from interactions of the CMB photons with the baryonic matter of the LSS, including the kinetic and thermal Sunyaev–Zel’dovich effects (kSZ and tSZ, respectively) \citep{kSZ_sunyaev_80,tSZ_sunyaev_72}, Rayleigh scattering by neutral hydrogen  \citep{basu_2004,lewis_rayleigh_13}, and fine-structure transitions in metals and ions \citep{basu_2004,chm_metals_reio,chm_metals_pol}.

Precision mapping of these anisotropies by COBE \citep{COBE1992_survey}, WMAP \citep{WMAP2003bennett_survey,WMAP2003spergel_survey}, and {\it Planck} \citep{Planck2014_survey,Planck2016_survey,Planck2020_survey} has enabled constraints at percent level on key cosmological parameters---including the baryonic and dark matter physical densities, and the amplitude and slope of the primordial power spectrum. Ground-based telescopes like the Atacama Cosmology Telescope (ACT, \citealt{ACT2011_survey,ACT2017_survey}) and the South Pole Telescope (SPT, \citealt{SPT2015_survey,SPT2019_survey}) have refined these measurements and upcoming ones such as the Simons Observatory (SO, \citealt{SimonsObs2019_survey})
 promise even higher accuracy.

At late times, the three-dimensional distribution of galaxies also carries cosmological information about the initial conditions of the Universe and its later evolution. Through the galaxy–halo connection, galaxies trace the underlying matter field with a certain bias, outlining the filamentary structure of the cosmic web. In this context, the study of cosmic voids has evolved significantly since their initial discovery. Thanks to their underdense nature, voids are less affected by the complex, non-linear processes that dominate overdense regions, making them particularly clean environments for probing cosmic expansion and dark energy \citep{
biswas2010_probe_de,bos2012_probe_de,pisani2015_vsf,pollina2016_probe_de}
, the properties of dark matter \citep{reed2015_probe_dm, baldi2018_probe_dm,arcari2022_probe_dm}, modified theories of gravity \citep{cai2015_probe_mg, barreira2015_probe_mg,cautun2018_probe_mg} and the presence of massive neutrinos \citep{massara2015_probe_n,  kreisch2019_probe_n, bayer2021_probe_n}. Their quiet interiors also shelter some of the most isolated galaxies, making them powerful laboratories for studying how environment shapes galaxy formation and evolution \citep{benson2003_probe_gal_ev, kreckel2012_probe_gal_ev, habouzit2020_probe_gal_ev}. The primary void observables are the void size function  \citep{sheth&vdweygaert2004_vsf, furlanetto2006_vsf, jennings2013_vsf,pisani2015_vsf, ronconi2017_vsf, contarini2019_vsf, contarini2023_vsf,verza2019_vsf, verza2024_vgcf_and_vsf_aprsd, verza2024_vsf, elena2025_vsf} and the void-galaxy cross-correlation function (with its distortions in redshift space) \citep{kaiser1987, alcock_1979, lavaux&wandelt2012_stacking_vgcf_aprsd,hamaus2015_vgcf_aprsd, nadathur2019_vgcf_aprsd, aubert2022_vgcf_rsd, correa2022_vgcf_aprsd, hamaus_2022_vgcf_aprsd, radinovic2023_vgcf_aprsd, degni2025_vgcf_aprsd}, which currently provide the tightest and most mature constraints from underdense regions.

Historically, cosmic voids have been treated as a byproduct of galaxy surveys and investigated mainly in wide-area datasets such as SDSS \citep{SDSS2000_survey,SDSS2009_survey}, VIPERS \citep{VIPERS2014_survey}, Pan-STARRS \citep{PanSTARSS2000_survey}, BOSS \citep{BOSS2013_survey}, and DES \citep{DES2005_survey,DES2016_survey,DES2022_survey}. Because of the multiscale nature of voids, which span tens to hundreds of megaparsecs, the ideal void survey would benefit from the combination of: large contiguous sky coverage to capture the rare, cosmologically sensitive large voids; dense galaxy sampling to resolve small-scale structures within voids; and preferably spectroscopic redshifts to minimize uncertainties in void–galaxy correlations. Moreover, a broad redshift coverage is also essential to track the evolution of void observables across cosmic time, which are particularly sensitive to dark energy and modified gravity---an aspect of growing importance given recent DESI hints of evolving dark energy \citep{DESI2025_BAO}. Projects such as the Wide-field Spectroscopic Telescope \citep[WST,][]{WST2024_survey} and the Stage-5 Spectroscopic Experiment \citep[Spec-S5,][]{SPEC52025_survey} aim to meet these demanding requirements but they remain long-term prospects. In the near future, significant progress will come from the synergy among modern surveys—DESI \citep{DESI2016_survey}, \emph{Euclid} \citep{EUCLID2011_survey}, SPHEREx \citep{SPHEREX2014_survey}, PFS \citep{PFS2014_survey}, LSST \citep{LSST2019_survey}, and the Nancy Grace Roman Space Telescope \citep{ROMAN2015_survey} which offer complementary perspectives and help mitigating systematic effects. In this work, the focus will be on \emph{Roman} ($1 \leq z \leq 3$), whose deep, high-resolution imaging and dense tracer sampling opens a new regime for void studies. Compared with \emph{Euclid} and DESI, \emph{Roman} combines finer angular detail with higher spectroscopic tracer density, which permits precise mapping of void structures at smaller scales. 

The study of voids is not isolated from that of the CMB. Early Void $\times$ CMB intensity cross-correlation studies focused on the ISW effect, with \cite{granett2008isw} reporting an anomalously strong signal by stacking supervoids and superclusters. While this result, in tension with $\Lambda$CDM, sparked extensive follow-up efforts \citep{papai2011isw, nadathur2012isw, flender2013isw, ilic2013isw, hernandezmonteagudo2013isw, aiola2015isw, hotchkiss2015isw, granett2015isw, cai2017isw, kovacs2018isw, kovacs2019isw, dong2021isw, hang2021isw, hansen2025isw}, it motivated a natural progression toward CMB lensing convergence ($\kappa$) maps, which offers a higher-significance and complementary probe. Subsequent studies focused on detecting this signal by stacking lensing maps on the angular position of thousands of voids, achieving significances from 3 to 17~$\sigma$ in recent years \citep{cai2017cmbk_isw, chantavat2016cmbk, raghunathan2020cmbk, vielzeuf2021DESY1cmbk, hang2021cmbk_isw, kovacs2022DESY3cmbk, vielzeuf2023demnuni_cmbk, camachociurana2024cmbk, demirbozan2024cmbk, sartori2024cmbk}.

Despite these successes, several challenges have prevented Void $\times$ CMB lensing from becoming a primary tool for cosmological parameter constraints. 
The main obstacle is the lack of an analytical model for the observable. The only theoretical framework was proposed by \citet{chantavat2016cmbk,chantavat2017cmbk} and relies on ideal conditions (multiple voids aligned along the line of sight). The rest of observational works dealing with real data and its complexities often aim to detect the signal and confirm its consistency with the $\Lambda$CDM model through a template-fitting method built upon the output of N-body cosmological simulations. This process involves comparing the observed signal to that derived from simulations by measuring one dimensional parameter $\rm A_{\kappa}=A_{data}/A_{model}$, which quantifies the amplitude required for the simulated convergence profile to match the observed lensing signal. In the absence of systematic differences between the data and the simulations $\rm A_{\kappa}$ is expected to be 
compatible with unity.

Deviations from unity have been reported in the literature, indicating mild tensions with the $\Lambda$CDM expectation up to 3$\sigma$ \citep{vielzeuf2021DESY1cmbk,hang2021cmbk_isw, kovacs2022DESY3cmbk,camachociurana2024cmbk} (see Figure \ref{fig:SoA_CMBxVoids} for a compilation of Void $\times$ CMB lensing analysis). Yet it remains unclear whether these discrepancies arise from genuine physical effects or from methodological artifacts introduced by the mock catalogs, void definitions, sample selection, or other aspects of the analysis. 

If such discrepancies hint at new physics, understanding them would require a detailed characterization of how the void lensing profile varies with cosmological parameters. This task is still limited by the scarcity of high-resolution, multi-cosmology simulations that coherently model both CMB and LSS observables. In general, simulations with CMB companions are already limited (e.g., Websky \citealp{Websky2020}, \textsc{Agora}\citealp{omori2022}, HalfDome \citealp{halfdome2025}), and most only model the standard $\Lambda$CDM cosmology\footnote{Some exceptions are DEMNUni \citep{DEMNUni2015Jcastorina,DEMNUni2016carbone} and the Gower Street simulations \citep{gowerstreet2025} but they lack the required resolution for the \emph{Roman} sample.}. This means that we can only compare observations to a very limited number of simulated scenarios, and typically only for the standard $\Lambda$CDM cosmology. At the same time, methodological choices can also impact the measured signal, an aspect that has never been systematically quantified. 

In this paper we focus on the latter point. We analyze how different mock catalogs and analysis choices affect the Void $\times$ CMB lensing profiles and their signal-to-noise ratios, while simultaneously forecasting the expected signal for the \emph{Roman} Space Telescope in combination with \emph{Planck}, Simons Observatory (SO), and CMB-S4–like experiments \citep{CMBS42016_survey} .

The paper is organized as follows. In Sect.~\ref{sec:datasets} we briefly describe the simulated datasets used for this work. Then, in Sect.~\ref{sec:observable}, we introduce the observable of interest. Section~\ref{sec:methodology}  presents the methodology. Section~\ref{sec:results} contains the results and discussions and Sect.~\ref{sec:conclusions} the conclusions.

\section{Simulated datasets}
\label{sec:datasets}
\subsection{Galaxy and void mock catalogs}
This work is conducted in the context of the Nancy Grace \emph{Roman} Space Telescope \citep{ROMAN2015_survey,zasowski2025roman}. \emph{Roman} is a NASA observatory scheduled for launch in late 2026, featuring a 2.4 m primary mirror---same as the Hubble Space Telescope---but with a field of view 100 times larger ($0.28$ deg$^2$) that will enable rapid, wide-area observations. The instrument combines visible/near-infrared imaging (0.48–2.3 $\mu$m) with slitless near-infrared spectroscopy (1.0–1.93 $\mu$m) to target cosmology, exoplanets, and infrared astrophysics. Its High Latitude Wide Area Survey (HLWAS) will map large-scale structure over $\sim$5100 deg$^2$, with spectroscopy covering $\sim$2415 deg$^2$ in its medium and deep tiers. The survey is expected to measure redshifts for roughly 14.2 million H$\alpha$, 3.6 million [OIII], and 1.3 million [OII] emission-line galaxies (ELGs) at $1\lesssim z\lesssim2.5$, producing a 3D map of the matter distribution about three times denser than that of \emph{Euclid} at $z=1.5$. 
 
All our datasets come from the \textsc{Agora} simulation\footnote{\url{https://yomori.github.io/Agora/index.html}}
 \citep{omori2022}, which provides a full-sky halo lightcone with self-consistent CMB secondary anisotropies maps (CMB lensing, tSZ/kSZ, CIB). In a companion paper \citep{perezsar2025_roman_mocks} we describe how we rely on \textsc{Agora} to build the mock galaxy and void catalogs employed here. As a summary, we use the \emph{Roman} mock catalog of \citet{zhai2021clustering} as a reference\footnote{We note that this catalog differs from what the Roman Time Allocation Committee (ROTAC) has recommended \citep{zasowski2025roman}: it's 20$\%$ smaller in  survey area (2000 vs. 2400 deg$^2$), and deeper (10$^{-16}$ vs. 1.5$\cdot 10^{-16}$ erg/s/cm$^2$). However, they are sufficiently similar for the present analysis.} and generate a suite of \emph{Roman}-like mocks for the \textsc{Agora} simulation \citep{omori2022} through an \emph{analog matching} algorithm. This algorithm links each \emph{Roman} galaxy with the most similar \textsc{Agora} halo counterpart in a multidimensional parameter space that includes halo mass, environment and other galaxy attributes. With this procedure we generate four families of mock catalogs with different levels of specification:

\begin{list}{-}{}
\item $N$ (number only): matches \emph{Roman}'s number density alone.
\item $NM$ (mass): matches number density and halo mass.
\item $NM,b_{\rm env}$ (mass + environment): additionally matches a local environment indicator defined as the density contrast in spheres of 5 $h^{-1}$~Mpc.
\item $NM,b_{\rm ELGs}$ (mass + galaxy-type indicators): includes mass and galaxy-type proxies (stellar mass, specific star formation rate), most closely resembling the semi-analytic model (SAM) mapping.
\end{list}

Each family is generated using four different selections of the extended \textsc{Agora} halo catalog to test the impact of different initial biases on the algorithm. The four variants are:

\begin{list}{-}{}
\item M-all: halos ranked by stellar mass
\item M-ELGs: halos ranked by stellar mass, restricted to ELG-host halos.
\item R-all: halos selected randomly.
\item R-ELGs: halos selected randomly, restricted to ELG-host halos.
\end{list}

Combining the four families with the four tracer selections we produce sixteen mock catalogs which span a range of specifications and biases relative to the \emph{Roman} ELG sample (See Appendix~\ref{sec:app_void_catalogs}, Figure \ref{fig:comparison_void_catalogs}).

The void catalogs are constructed using both 3D and 2D void-finders. The 3D catalog is generated with \texttt{REVOLVER}\footnote{\url{https://github.com/seshnadathur/Revolver}}
 (REal-space VOid Locations from surVEy Reconstruction), based on a modified version of the \texttt{ZOBOV} algorithm \citep{neyrinck2008zobov}. The 2D catalog is obtained as described in \cite{sanchez20162d}, adopting smoothing scales of ${\rm sm_{VF}} = 10~h^{-1}~\mathrm{Mpc}$ and ${\rm sm_{VF}} = 5~h^{-1}~\mathrm{Mpc}$, and a slice thickness of $s \approx 70~h^{-1}~\mathrm{Mpc}$. Further details on the void finders and about the parameter choices can be found in \cite{perezsar2025_roman_mocks}. Figure \ref{fig:comparison_void_catalogs} in the appendix summarizes the main properties of each void catalog.

\begin{table*}[]
\caption{Summary of the main specifications of the LSS survey and CMB experiments considered in this work.}
\hspace*{-0.2cm} 
\resizebox{1.03\textwidth}{!}{
\begin{tabular}{llllllll}
\hline
\multicolumn{1}{|l}{} & \multicolumn{1}{l|}{} &  &  &  &  &  & \multicolumn{1}{l|}{} \\
\multicolumn{2}{|c|}{\textbf{LSS survey}} & \multicolumn{1}{c}{\textbf{Duration}} & \multicolumn{1}{c}{\textbf{Final area}} & \multicolumn{1}{c}{\textbf{\begin{tabular}[c]{@{}c@{}}Wavelengths\\ {[}$\mu$m{]}\end{tabular}}} & \multicolumn{1}{c}{\textbf{\begin{tabular}[c]{@{}c@{}}Angular \\ resolution\\ {[}arcsec{]}\end{tabular}}} & \multicolumn{1}{c}{\textbf{Detectors}} & \multicolumn{1}{c|}{\textbf{Sensitivity}} \\
\multicolumn{1}{|l}{} & \multicolumn{1}{l|}{} &  &  &  &  &  & \multicolumn{1}{l|}{} \\ \hline
\multicolumn{1}{|l}{} & \multicolumn{1}{l|}{} &  &  &  &  &  & \multicolumn{1}{l|}{} \\
\multicolumn{1}{|c}{\textbf{Roman}} & \multicolumn{1}{l|}{\begin{tabular}[c]{@{}l@{}}Space (L2)\\ 2.4m mirror and two instruments:\\ \\ 1) Wide Field Instrument (WFI)\\     - For imaging: Filter wheel with 7 \\        standard filters and 1 wide filter \\        (H-band)\\     - For spectroscopy: prism (R$\sim$100) \\       for supernovae and other transients\\       science and grism (R $\sim$435-865) \\       for the HLWAS.\\ \\ 2) Coronograph\end{tabular}} & \multicolumn{1}{c}{\begin{tabular}[c]{@{}c@{}}2026-2031\end{tabular}} & \begin{tabular}[c]{@{}l@{}}$\sim$12\% of the sky\\               $\sim$6\% (spectroscopy)\\ \\ Total HLWAS area: \\  5117 deg$^2$\\   - Deep tier: 19.2 deg$^2$\\      7 imaging bands\\   - Medium tier: 2415 deg$^2$\\      3 imaging bands + grism\\   - Wide Tier: 2702 deg$^2$ \\     using H-band imaging only\end{tabular} & \multicolumn{1}{c}{\begin{tabular}[c]{@{}c@{}}Imaging 0.48–2.3 \\ Spectroscopy 1.0–1.93\end{tabular}} & \multicolumn{1}{c}{$\sim$0.1} & \multicolumn{1}{c}{18} & \multicolumn{1}{c|}{\begin{tabular}[c]{@{}c@{}}5$\sigma$ point-source depth\\  of $\approx$\\ 26.2 to 28.2 mag\\  (filter dependent)\end{tabular}} \\
\multicolumn{1}{|l}{} & \multicolumn{1}{l|}{} &  &  &  &  &  & \multicolumn{1}{l|}{} \\ \hline \noalign{\vskip 0.15cm} \hline
\multicolumn{1}{|l}{} & \multicolumn{1}{l|}{} &  &  &  &  &  & \multicolumn{1}{l|}{} \\
\multicolumn{2}{|c|}{\textbf{CMB surveys}} & \multicolumn{1}{c}{\textbf{Duration}} & \multicolumn{1}{c}{\textbf{Final area}} & \multicolumn{1}{c}{\textbf{\begin{tabular}[c]{@{}c@{}}Frequency \\ (GHz) \\ and Bands\end{tabular}}} & \multicolumn{1}{c}{\textbf{\begin{tabular}[c]{@{}c@{}}Angular \\ resolution\\ {[}arcmin{]}\end{tabular}}} & \multicolumn{1}{c}{\textbf{Detectors}} & \multicolumn{1}{c|}{\textbf{\begin{tabular}[c]{@{}c@{}}Sensitivity \\ (co-added)\\ {[}$\mu$K $\cdot$ arcmin{]}\end{tabular}}} \\
\multicolumn{1}{|l}{} & \multicolumn{1}{l|}{} &  &  &  &  &  & \multicolumn{1}{l|}{} \\ \hline
\multicolumn{1}{|l}{} & \multicolumn{1}{l|}{} &  &  &  &  &  & \multicolumn{1}{l|}{} \\
\multicolumn{1}{|c}{\textbf{Planck}} & \multicolumn{1}{l|}{\begin{tabular}[c]{@{}l@{}}Space (L2)\\ 1.5 m mirror with 2 main instruments:\\ Low Frequency Instrument (LFI) \\ High Frequency Instrument (HFI)\end{tabular}} & \multicolumn{1}{c}{\begin{tabular}[c]{@{}c@{}}2009 - 2013, \\ completed\end{tabular}} & \multicolumn{1}{c}{full sky} & \multicolumn{1}{c}{30 - 850 (9 bands)} & \multicolumn{1}{c}{\begin{tabular}[c]{@{}c@{}}33 to 5\\ (freq.\\ dependent)\end{tabular}} & \multicolumn{1}{c}{74} & \multicolumn{1}{c|}{26 - 28} \\
\multicolumn{1}{|l}{} & \multicolumn{1}{l|}{} &  &  &  &  &  & \multicolumn{1}{l|}{} \\ \hline
\multicolumn{1}{|l}{} & \multicolumn{1}{l|}{} &  &  &  &  &  & \multicolumn{1}{l|}{} \\
\multicolumn{1}{|c}{\textbf{\begin{tabular}[c]{@{}c@{}}Simons\\ Observatory\\ (SO)\end{tabular}}} & \multicolumn{1}{l|}{\begin{tabular}[c]{@{}l@{}}Ground (Cerro Toco, Chile)\\ 1x5m Large Aperture Telescope (LAT) \\ 3x0.5m Small Aperture Telescopes (SATs)\end{tabular}} & \multicolumn{1}{c}{\begin{tabular}[c]{@{}c@{}}2025-2034, \\ ongoing\end{tabular}} & \multicolumn{1}{c}{$\sim$61\% of sky} & \multicolumn{1}{c}{27 - 280 (6 bands)} & \multicolumn{1}{c}{7.4 to 0.9} & \multicolumn{1}{c}{$\sim$60000} & \multicolumn{1}{c|}{2.6 - 2.8} \\
\multicolumn{1}{|l}{} & \multicolumn{1}{l|}{} &  &  &  &  &  & \multicolumn{1}{l|}{} \\ \hline
\multicolumn{1}{|l}{} & \multicolumn{1}{l|}{} &  &  &  &  &  & \multicolumn{1}{l|}{} \\
\multicolumn{1}{|c}{\textbf{CMB-S4-like}} & \multicolumn{1}{l|}{\begin{tabular}[c]{@{}l@{}}Ground (South Pole + Chile) \\ Mutiple 6m LATs and 0.5m SATs\end{tabular}} & \multicolumn{1}{c}{-} & \multicolumn{1}{c}{\begin{tabular}[c]{@{}c@{}}$\sim$70\% of the sky\\ (Legacy Survey) \\ $\sim$3\% (ultra deep)\end{tabular}} & \multicolumn{1}{c}{$\sim$30 - 300 (11 bands)} & \multicolumn{1}{c}{\begin{tabular}[c]{@{}c@{}}30 \\ (large scales)\\ 1-1.5 \\ (high res)\end{tabular}} & \multicolumn{1}{c}{\textgreater 500000} & \multicolumn{1}{c|}{$\sim$1} \\
\multicolumn{1}{|l}{} & \multicolumn{1}{l|}{} &  &  &  &  &  & \multicolumn{1}{l|}{} \\ \hline
\end{tabular}}
\label{tab:lss_cmb_specs}
\end{table*}

\subsection{CMB $\kappa$ maps}
The CMB convergence ($\kappa$) map is taken from \textsc{Agora} as well. Details on how this map is created can be found either on \citealp{perezsar2025_roman_mocks} or \citealt{omori2022}. To simulate realistic survey conditions, we generate noise realizations from the noise power spectra of each CMB survey and add them to the noise-free \textsc{Agora} CMB $\kappa$ map.\\

We use \emph{Planck} as the baseline for the current generation of experiments, while SO and a CMB-S4-like serve as future reference cases. All three fully overlap with the Roman footprint. Although the original CMB-S4 project is inactive, we still use a CMB-S4-like setup as an optimistic, ultra-low-noise reference.\\

Before the Voids $\times$ CMB $\kappa$ cross-correlation, we pre-process the map by:

\begin{list}{}{}
\item (i) degrading the HEALPix\footnote{HEALPix URL site: \url{https://healpix.sourceforge.io/}} \citep{healpix} resolution from $N_{\rm side}=8192$ to $N_{\rm side}=512$, since the angular scales probed by voids are sufficiently large that a higher-resolution pixelization does not provide additional information;
\item (ii) scaling values by $10^3$ for convenient units;
\item (iii) applying Gaussian smoothing when needed to reduce small-scale noise;
\item (iv) remove large-scale patterns introduced by incomplete sky coverage or anisotropic noise.
\item (v) masking the sky to match the 2000 deg$²$ \emph{Roman}-like footprint.
\end{list}

The instrumental specifications for the LSS and CMB surveys used in this work are summarized in Table \ref{tab:lss_cmb_specs}.

\section{Observable: ELG-traced voids and CMB lensing convergence maps}
\label{sec:observable}
\subsection{Voids traced by ELGs}

Cosmic voids are not direct observables since their existence is inferred from the distribution of galaxies (or other tracers). Thus, any void catalog is a biased tracer of the underlying LSS. Their properties, including size, density profiles and number, depend not only on the specific void finding algorithm but also on the tracer's number density, selection function, and bias. 

In this work we use Emission Line Galaxies (ELGs) as tracers, following the \emph{Roman} survey strategy. ELGs are galaxies characterized by strong emission lines from ionized gas, primarly arising from active star formation and, to a lesser extent, by active galactic nuclei (AGN) \citep{gonzalez2020}. Their optical and infrared nebular lines such as [OII] ($\lambda$3726, $\lambda$3729~\AA), H$\alpha$ ($\mathrm{\lambda}$ 6563~\AA) and [OIII] ($\lambda$4959, $\lambda$5007~\AA) allow precise spectroscopy at $0.5 \lesssim z \lesssim 2$ \citep{drinkwater2010, comparat2013}. Because the line-strength hierarchy (H$\alpha$ $>$ [OIII] $>$ [OII]) governs redshift robustness, \emph{Roman} relies on H$\alpha$ and [OIII] for $1 < z < 2$, and on [OIII] and [OII] for $2 < z < 3$, while weaker emission features primarily contribute to interloper contamination through redshift misidentification.

Observations and models show that ELGs reside in lower-density environments compared to more massive tracers like Luminous Red Galaxies (LRGs). Specifically, ELGs are found primarily in filaments ($\sim$50$\%$) and sheets ($\sim$30$\%$). They typically inhabit lower-mass dark matter halos, with characteristic halo masses of order $M_{\rm h}\sim10^{11.5}-10^{12}~h^{-1}\rm{M_\odot}$. Furthermore, their halo occupation strongly depends on environment: centrals dominate in voids and sheets, while satellites become more common in filaments and knots (for ELG–halo connection studies see \cite{favole2016,gonzalez2020,Hadzhiyska2021,rocher2023,yu2024,yuan2025}). Environmental and assembly bias effects, though generally weak for ELGs, should therefore be considered \citep{gonzalez2020}.

\subsection{CMB lensing convergence $\kappa$}
The CMB lensing convergence, denoted as $\kappa$,
is related to the deflection angle $\alpha$ experienced by CMB photons as they travel towards the observer from the last scattering surface at $z\sim 1100$. Under the Born approximation\footnote{The Born approximation assumes that the gravitational potential of the lens only causes a small perturbation to the light path, allowing us to treat it as a straight line.} it reads:

\begin{equation}
\vec{\alpha} (\vec{\theta} ) = \nabla_\Omega \biggl[ \frac{2}{c^2} \int_0^{\chi_{\rm LS}} d\chi \frac{\chi_{\rm LS}-\chi}{\chi_{\rm LS}\chi} \Phi (\chi,\vec{\theta})\biggr] \equiv \nabla_\Omega \phi (\vec{\theta}),
\label{eq:lenseq1}
\end{equation}
with $\phi$ is the lensing potential, defined as the projection of the 3D gravitational potential $\Phi$. The operator $\nabla_\Omega$ denotes the angular gradient on the unit sphere defined as 
$\nabla_\Omega = \hat{\theta}\,\frac{\partial}{\partial \theta} + \frac{\hat{\varphi}}{\sin\theta}\,\frac{\partial}{\partial \varphi}$, 
$\chi$ denotes comoving distances, and $\chi_{\rm LS}$ the distance to the CMB last scattering surface. 

The CMB along direction $\vec{\theta}$ is remapped to $\vec{\theta}^{\rm obs}\simeq \vec{\theta} + \nabla_\Omega \phi(\vec{\theta})$. The measurable lensing information resides in the spatial derivatives of $\alpha$, which describe how extended CMB regions are sheared and magnified. Inversion methods recover these derivatives through the magnification (Jacobian) matrix:

\begin{equation}
A_{ij} \equiv \frac{\partial \theta_i^{\rm obs}}{\partial \theta_j} = \delta_{ij} + \frac{\partial \alpha_j}{\partial \theta_i} =
\begin{pmatrix}
1 - \kappa - \gamma_1 & -\gamma_2 + \omega \\
-\gamma_2 - \omega & 1 - \kappa + \gamma_1
\end{pmatrix}.
\label{mag_def}
\end{equation}
This matrix decomposes into three components:
\begin{itemize}
    \item \textbf{Convergence} $\kappa$: Isotropic magnification or demagnification of a sky patch. Positive convergence ($\kappa>0$) occurs in overdense regions, slightly compressing the apparent angular scale of CMB fluctuations. Negative convergence ($\kappa<0$) occurs in underdense regions (voids), slightly stretching the CMB patches.
    \item \textbf{Shear} $\gamma_1 + i\gamma_2$: Anisotropic streching of a circular sky patch into an elliptical one without changing the area.
    \item \textbf{Rotation} $\omega$: antisymmetric distortion corresponding to local rotations. In the Born approximation it cancels out.
\end{itemize}

Mathematically and still under the Born approximation, the convergence can be defined as half the (angular) Laplacian of the lensing potential: 
\[\kappa(\vnh) = \frac{1}{2} \Delta_{\Omega} \phi(\vnh) = \frac{1}{c^2} \int_0^{\chi_{\rm LS}} d\chi \,\frac{\chi_{\rm LS} - \chi}{\chi_{\rm LS} \chi} \nabla_\Omega^2 \Phi(\chi,\vnh)
\]

\begin{equation}
\phantom{xxxxxxxxxxxxxxx}
 \simeq \,\frac{3 H_0^2 \Omega_m}{2 c^2} \int_0^{\chi_{\rm LS}}d\chi\, \frac{\chi_{\rm LS} - \chi}{\chi_{\rm LS} a(\chi)} \,\chi\,\delta(\chi,\vnh).
 \label{eq:kappa1}
\end{equation}

In the last equation we have used the Poisson equation in comoving units and converted the spatial Laplacian ($\Delta$) into the angular one ($\Delta_\Omega$) via the approximation $\Delta \simeq 1/\chi^2 \Delta_{\Omega} = 1/\chi^2 \nabla_\Omega^2$:
\begin{equation}
\nabla_\Omega^2 \Phi \simeq \chi^2\frac{3 H_0^2 \Omega_m}{2 a} \delta,
\label{eq:poisson1}
\end{equation}

where the line-of-sight contribution to the spatial Laplacian has been neglected, $a(\chi)$ denotes the cosmic scale factor, $\delta$ corresponds to the matter density contrast, and $H_0$ and $\Omega_m$ refer to the Hubble constant and the matter density parameter, respectively. According to Eq.~\ref{eq:kappa1} the convergence essentially maps the (distance-weighted) matter density contrast along the line of sight.

\section{Methodology}
\label{sec:methodology}

The template-fitting method is the summary statistic commonly used to study the lensing imprint of voids \citep{raghunathan2020cmbk, kovacs2022DESY3cmbk, sartori2024cmbk}. This technique consists in extracting a $\kappa$ profile from simulations as a reference template and comparing it to the real observations.
The comparison is quantified by the amplitude parameter $A_{\kappa}$, defined as the ratio between the observed and simulated CMB $\kappa$ profiles at common angular scales. Its signal-to-noise ratio, $\rm{S/N} = A_{\kappa}^{\rm ML} / \sigma_{A_{\kappa}}$, is obtained from the maximum-likelihood estimate $A_{\kappa}^{\rm ML}$ and its uncertainty $\sigma_{A_{\kappa}}$, derived from the likelihood width.

The accuracy of the template fitting method depends on several factors like the assumed cosmology, and how closely the mock galaxies trace the spatial distribution of the real data. This effective 1-parameter $A_{\rm \kappa}$ fit, while a standard of the field, introduces a model dependence, potentially obscuring more complex or subtle deviations present in real data by compressing them into an overly simplistic representation. In this work, we do examine how different mock catalogs can influence the profiles used to compute $A_{\rm \kappa}$; however revising the method is beyond the scope of this work.

\subsection{Estimating the $\kappa$ profiles}

Once we have our void catalog and the CMB $\kappa$ map we perform their cross-correlation. For this we stack the $\kappa$ signal at the positions of an ensemble of voids since the S/N of a single void is much smaller than unity \citep{krause2013wl}. This stacking can be done via two different approaches: 

\begin{itemize}
    \item {\it Profile-based}. For each void, we compute the radial $\kappa$ profile around its center in angular bins up to a maximum radius. We adopt either $\theta_{\rm max}=5R_v$ or $\theta_{\rm max}=5^\circ$. In the former case, profiles are rescaled by the void angular radius $R_v$ before stacking; in the latter, no rescaling is applied.

    \item {\it Image-based}. We stack $\kappa$ map cutouts centered on each void, rescaled to a fixed grid of $200\times200$ pixels spanning $10R_v$. All cutouts are brought to this common scale by changing its resolution prior to stacking. Once we have the final image, we obtain the profile. 
    
\end{itemize}

We then sum all profiles/patches and measure the average $\kappa$ signal in different concentric radius bins around the void centers. We use 25 bins of $\Delta = 0.2$ (in $r/R_{\mathrm{v}}$ or degrees, depending on whether rescaling is applied). We note that, while the second approach offers a direct image of the signal, it is computationally demanding, especially for large void samples.

\subsection{Estimating associated errors}

We test the null hypothesis (i.e, the probability that the predicted signal is given by a random fluctuation) by stacking 1000 random CMB $\kappa$ realizations at the void positions. With this we aim to quantify the two main sources of noise that affect the void-lensing cross-correlation: i) the instrumental noise, and ii) the intrinsic noise of the $\kappa$ map. The latter is given by the contribution to the void $\kappa$ signal from structures (voids and clusters) that either lie along the line of sight or in the surroundings of the target voids, and are counted multiple times due to void overlapping. These contributions are orders of magnitude larger than the signal of interest, and this inevitably results in an intrinsic scatter on the $\kappa$ measurement.

We assume both components are Gaussian and model them through their angular power spectra. The random maps are generated with \texttt{healpy/synfast} at $N_{\rm side}=512$ and $\ell_{\rm max}=3N_{\rm side}-1$. The intrinsic $\kappa$-noise component uses the power spectrum measured from the \textsc{Agora} $\kappa$ map, while instrumental noise is added using the noise power spectra of each experiment. Both components are produced separately, $\kappa_{\rm random} = \kappa^{\text{AGORA,random}} + \kappa^{\text{Inst Noise}}$, allowing us to switch them on or off and explore noise schemes ranging from noise-free (intrinsic scatter only) to realistic configurations including instrumental noise from \emph{Planck}, SO, or CMB-S4-like surveys. Finally, we also apply the \emph{Roman} mask to account for the sample variance associated to a finite sky coverage. 

For each void, we store the $\kappa$ profile from the \textsc{Agora} lensing map and from all random realizations. In this way we can later study sub-samples of void populations under different observational configurations without recomputing the stacks.

\begin{figure*}[h]
    \centering
    \includegraphics[width=13cm]{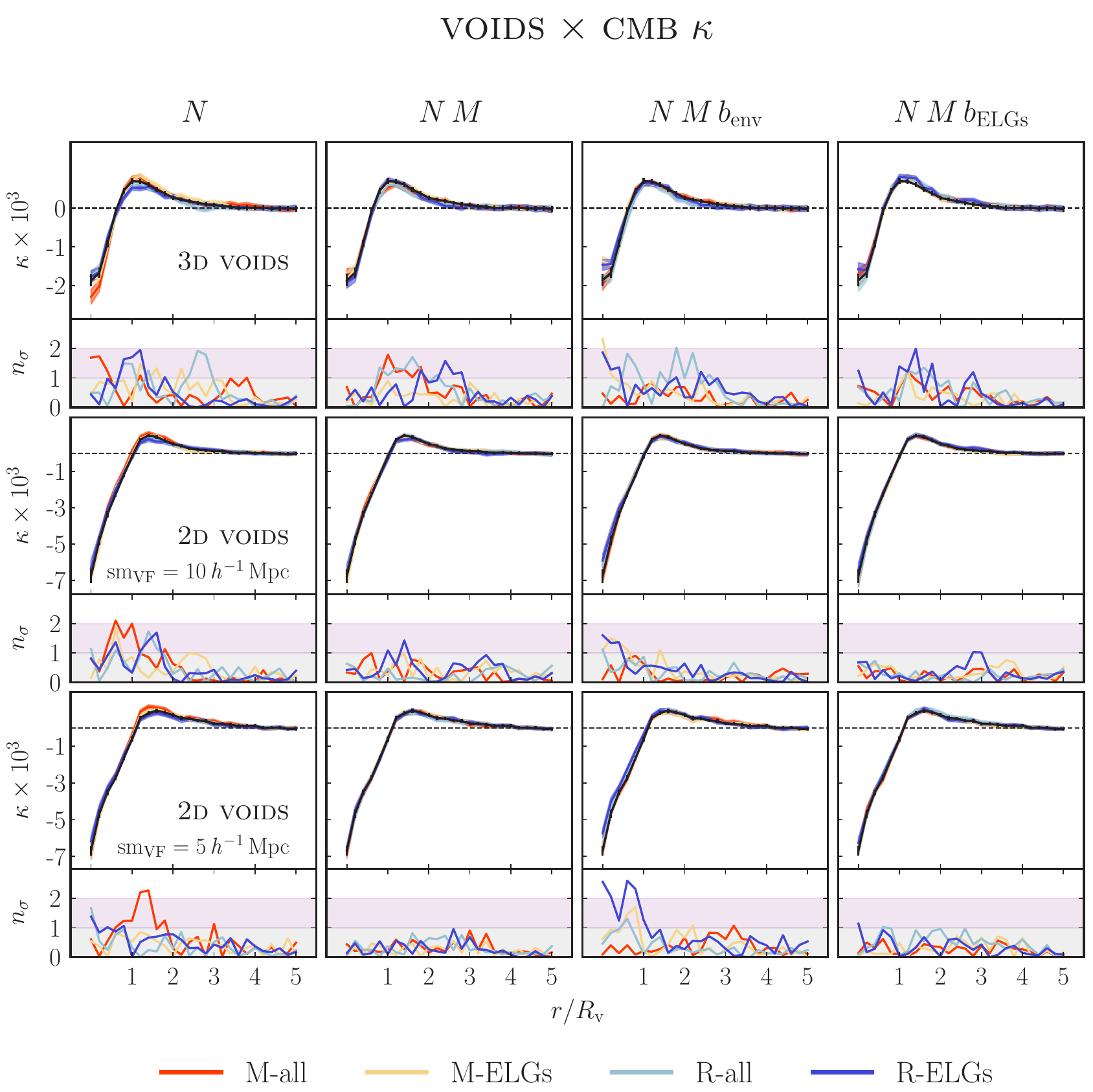} 
    \caption{Void $\times$ CMB lensing profiles for the different mock catalogs. Columns correspond to different mock families; colors indicate tracer bias. The mean profile we use for comparison is computed as the average over the mocks that are consistent with the one- and two-point statistics of the \emph{Roman} reference catalog. Residuals show deviations (in $\sigma$) from the mean profile and are computed as the difference between each profile and the mean, normalized by the square root of their uncertainties.}
    \label{fig:lensing_comparison_main}
\end{figure*}

\subsection{The signal-to-noise (S/N) ratio}

We estimate the signal-to-noise ratio using a likelihood approach to incorporate the full data vector and its covariance. Simpler estimators---such as the ratio of mean signal to mean noise or the peak anisotropy---neglect correlations between radial bins and the scale dependence of the noise. The process consist of maximizing the likelihood by fitting the observed profile $\kappa^{\rm OBS}$ to a simulated template $\kappa^{\rm SIM}$. In this work, we do not have observed data but rather synthetic profiles from \textsc{Agora}, so we set $\kappa^{\rm OBS}$ = $\kappa^{\rm SIM}$ (corresponding to a maximum-likelihood amplitude of $A_{\kappa}^{\rm ML}=1$) and derived the expected S/N from the likelihood width. The width is quantified as the $68\%$ confidence interval of the cumulative distribution (16th–84th percentiles), corresponding to $\pm1\sigma$ for Gaussian-like posteriors\footnote{Given that our intrinsic $\kappa$ signal and noise mocks are Gaussian, the resulting PDF for $A_\kappa$ is also expected to be Gaussian.}. 

The likelihood is defined as:

\begin{eqnarray} 
{\cal L}(A_{\kappa}) & \propto  &\exp{\biggl[ \frac{-\chi^2(A_{\kappa})}{2}\biggr]},\\
{\chi}^2 (A_{\kappa}) &= & \sum_{ij} \left(\kappa_{i}^{{\rm OBS}}-A_{\rm \kappa}\cdot\kappa_{i}^{\rm SIM}\right) C_{ij}^{-1} \left(\kappa_{j}^{{\rm OBS}}-A_{\rm \kappa}\cdot\kappa_{j}^{\rm SIM}\right),
\label{eq:likel1}
\end{eqnarray}
with the $i,j$ indexes running through angular bins and $C_{ij}$ being the covariance matrix:
    \begin{equation}
        C_{ij} = \frac{1}{N-1} \sum_n^N (\kappa_i^n - \bar{\kappa}_i)(\kappa_j^n - \bar{\kappa}_j) \, ,
    \end{equation}

Here $N=1000$ is the number of random CMB-$\kappa$ realizations; $\kappa_i^n$ is the convergence in the $i$-th radial bin for the $n$-th realization, and $\bar{\kappa}_i$ is the mean over all realizations. We correct the inverse covariance using the Hartlap factor \citep{hartlap2007}, given by
$C^{-1}_{ij}=(N_{\rm randoms}-N_{\rm bins}-1)/(N_{\rm randoms}-1)\langle C^{-1}_{ij}\rangle$. This Hartlap factor compensates for biases arising from inverting a covariance matrix estimated from a limited number of realizations.

\subsection{Methodological considerations}

The main limitation of this work is the use of a single sky patch, despite the availability of a full-sky lightcone. Ideally, multiple independent realizations would better sample cosmic variance and strengthen the robustness of the results. In practice, however, computing the cross-correlation signal and its covariance over many realizations is computationally and storage prohibitive within our resources.

We justify this single-patch approach on two fronts. First, for our methodological comparison, the single realization serves as a controlled environment in which we apply all 16 mock types and stacking techniques to the same large-scale structure (LSS), ensuring that any observed variations in S/N or profile behavior are driven strictly by the methods themselves. Second, for our lensing forecast, the variance is dominated by CMB noise (detector noise and primary CMB fluctuations) rather than the LSS distribution. As shown in previous work \citep{sartori2024cmbk}, sample variance contributes only a small fraction of the total covariance compared to the noise level in current \emph{Planck} maps.

\section{Results}
\label{sec:results}

We present the first forecast of the Voids $\times$ CMB $\rm \kappa$ signal for the \emph{Roman} telescope. In the process, we investigate how mock catalogs of varying levels of complexity affect the signal and its signal-to-noise ratio for different void types, void finders and other methodological choices. We then incorporate instrumental noise and report signal-to-noise (S/N) ratios for various background noise scenarios, including {\it Planck}, SO, and S4-type experiments.

\begin{figure*}[!]
    \centering
    \includegraphics[width=15cm]{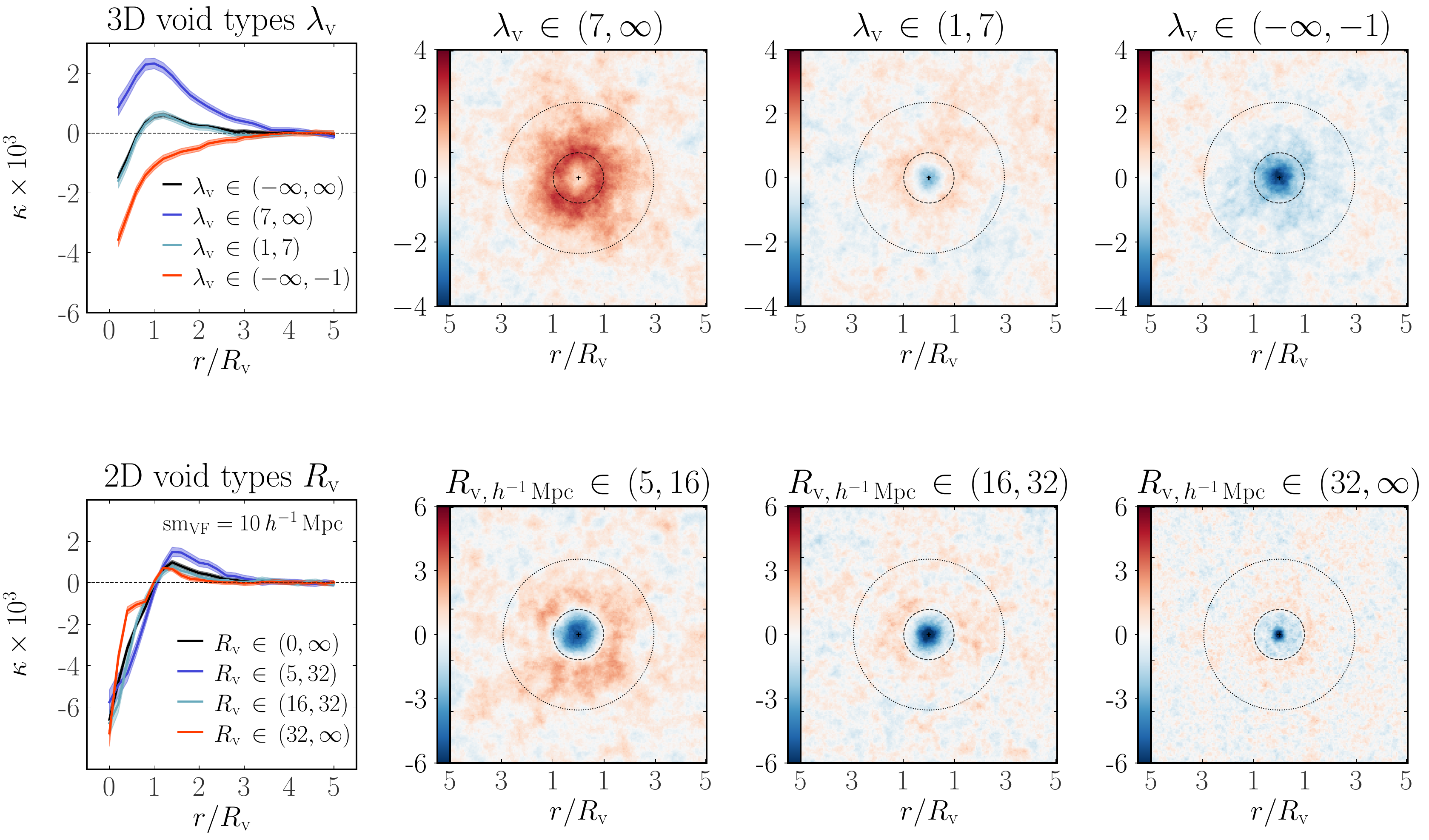} 
    \caption{Void $\times$ CMB lensing profiles and corresponding footprints for the specific mock catalog $N\,M$ (M-all) displaying the different void populations for both 3D and 2D samples. In the left panels, the profile corresponding to the average through all voids is depicted in black, and falls close to the intermediate bins in both $\lambda_v$ and $R_{\rm v}$, depicted in green.}
\label{fig:footprints_and_lensing_by_types}
\end{figure*}

\subsection{Impact of mock catalogs on void $\times$ CMB $\rm \kappa$ signal}
\label{sec:results_impact_mocks_in_lensing_signal}

We quantify the impact of the mock catalog choice by comparing the Void $\times$ CMB lensing signal measured from each mock to a reference mean lensing profile. This reference profile is constructed by averaging the lensing signal from the mock catalogs that successfully reproduce the void and galaxy one- and two-point statistics of the \emph{Roman} reference catalog \citep{perezsar2025_roman_mocks} \footnote{These were the cases of  $N\,M$ and $N\,M\,b_{\rm ELGs}$ mock families, regardless of the input tracer selection and $N\,M\,b_{\rm env}$ (M-all).}. From now on we will refer to those as \emph{Roman-matched mocks} \footnote{These catalogs are available and can be provided under request to the first author to perform further forecasting efforts for \emph{Roman}.}. Note that the \emph{Roman} reference catalog itself is not used because no associated CMB lensing maps are available \citep{zhai2021clustering}. 

As we have mentioned, cosmic voids act as divergent lenses, producing a de-magnification signal, $\kappa<0$, opposite to the convergent lensing of clusters, $\kappa>0$. The lensing amplitude scales with the density contrast, with deeper voids generating stronger effects. For 2D voids, the void radius, $R_{\rm v}$, is defined by the first crossing of the mean density (the zero-crossing of $\kappa$), 
while for 3D voids $R_{\rm v}$ aligns more closely with the compensation wall if present. The compensation wall is an overdense shell surrounding the void where the density rises above the cosmic mean and sometimes compensates for its central underdensity.

\subsubsection{Total sample}
Figure \ref{fig:lensing_comparison_main} presents the Void $\times$ CMB $\kappa$ signal when all voids in the sample are considered. There we can see that the significant differences observed in the one- and two-point statistics across mock catalogs (see \citealt{perezsar2025_roman_mocks}) do not translate into major differences in the lensing profile. Deviations remain below $2\sigma$ relative to the mean profile and would be indistinguishable once instrumental noise (from surveys such as \emph{Planck} and SO) is included. See Appendix~\ref{sec:app_signal_with_IN}, Figure \ref{fig:signalk_with_different_IN}.

For 3D voids, variations are slightly larger due to the impact of intervening large-scale structure and the limited number of detected voids in the \emph{Roman}-like area. In contrast, 2D voids are defined from a projected density field, therefore part of the same foreground and background structure that contributes to the lensing signal also contributes to defining the void itself, producing more stable profiles. 

When a smaller smoothing is used for the 2D void finder (see 2D voids with ${\rm sm_{VF}} =  5~h^{-1}~$Mpc) the mocks that better reproduce the reference catalog (namely $N\,M$ and $N\,M\,b_{\rm ELGs}$) show reduced scatter, indicating an increased sensitivity to the mock catalog. Nonetheless, all Void $\times$ CMB lensing profiles remain consistent in shape within the errors below 2$\sigma$. This consistency would only increase once the CMB $\kappa$ instrumental noise is included. The lack of significant deviations indicates that differences between the mocks have too small of an impact on the Void $\times$ CMB lensing signal to generate a spurious tension with $\Lambda$CDM if we consider the instrumental noise of \emph{Planck} or SO.

In the Appendix \Cref{fig:lensing_rescaled_and_not_rescaled_with_Rcuts_3d,fig:lensing_rescaled_and_not_rescaled_with_Rcuts_2dsm10,fig:lensing_rescaled_and_not_rescaled_with_Rcuts_2dsm5}, we further explore the effect of applying radius cuts to the void samples (e.g., selecting voids with sizes above and below twice the mean separation particle, mps). This allows us to both test the impact of excluding smaller, less reliable voids likely impacted by shot noise, and also to examine the noise properties of the resulting subsets. The overall conclusions remain consistent with the full sample, though uncertainties increase due to reduced statistics.

\begin{figure*}[!b]
    \centering
    \hspace*{-0.6cm} 
    \includegraphics[width=19.5cm]{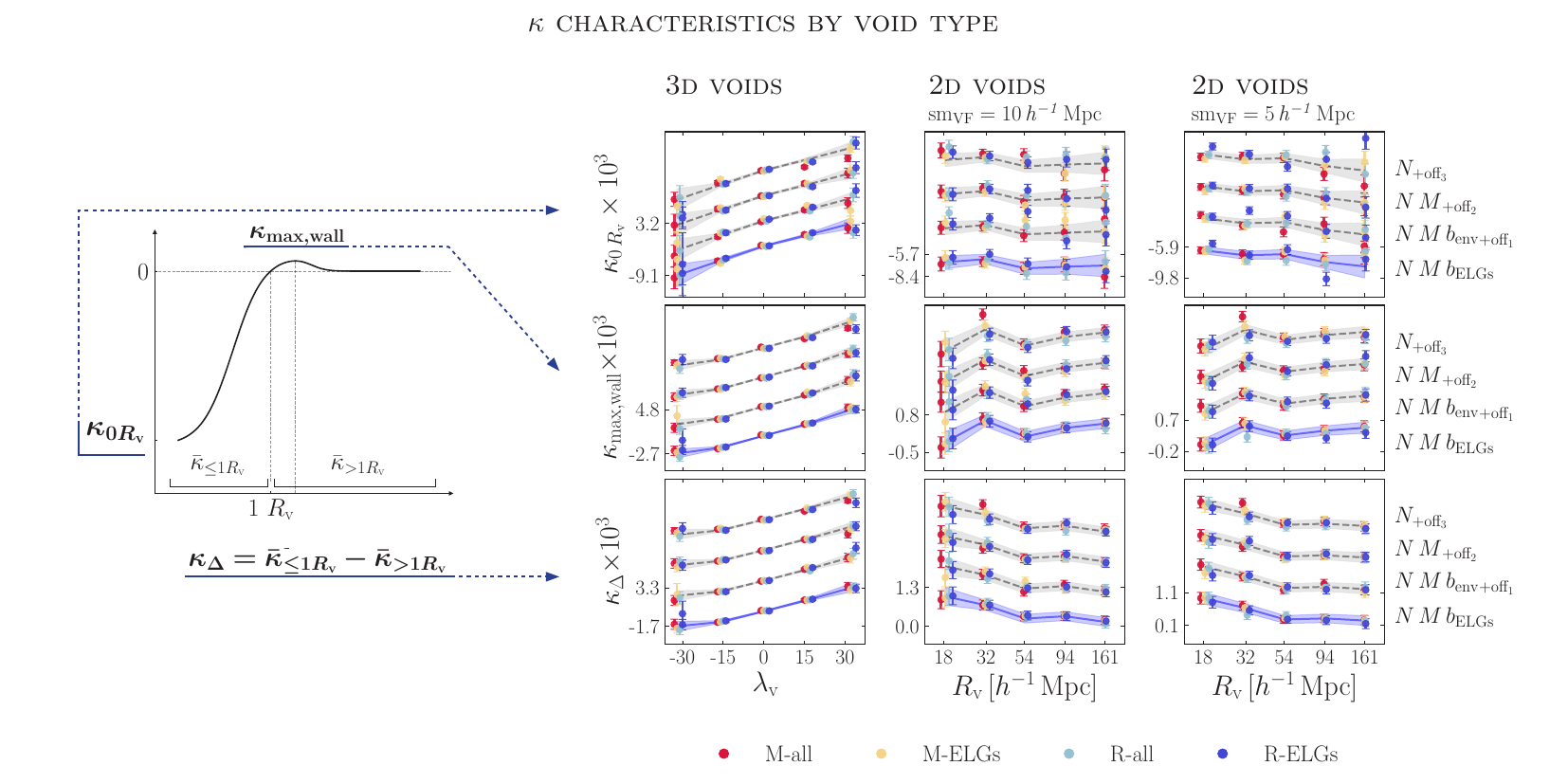} 
    \caption{Distribution of lensing-profile parameters ($\rm \kappa_{0R_{\rm v}}$, $\rm \kappa_{max,wall}$, $\rm \kappa_\Delta$) for different void types in both 3D and 2D catalogs. The blue curve shows the real scale profile, while the other curves, corresponding to different mock catalog types, are offset along the $y$-axis for clarity. Colors indicate different tracer biases. All profiles are compared to the mean profile of the \emph{Roman-matched mocks} that successfully reproduce the one- and two-point statistics of the \emph{Roman} Reference Catalog.}
    \label{fig:kcharact_rescaled}
\end{figure*}

\subsubsection{Splitting by type}

The apparent consistency across mock catalogs and bias selections in the Void $\times$ CMB lensing signal may partly arise from averaging over the entire sample of voids which can mask differences between void populations. This is a common concern for any void measurement.

There are fundamentally two main void classes: void-in-clouds \citep{sheth&vdweygaert2004_vsf}, which are surrounded by a large overdensity and experience compression and a slower expansion due to the inward gravitational pull; and void-in-voids, which are embedded within larger underdense regions and expand freely with the cosmic expansion. For 3D voids, we use the parameter $\lambda_v$ \citep{nadathur2017lambda}, defined as $\lambda_v = \bar{\delta}_{\rm avg} \left(R_{\rm v}/[h^{-1}\mathrm{Mpc}]\right)^{1.2}$ \footnote{$\bar{\delta}_{\rm avg}$ is the average density contrast within the void and $R_{\rm v}$, the void radius.}, to distinguish between  classes. Negative $\lambda_v$ corresponds to underdense void-in-voids, positive $\lambda_v$ indicates void-in-clouds with a strong surrounding compensation wall. $\lambda_v$ values near zero correspond to intermediate cases. In 2D, we use the void radius $R_{\rm v}$ as a proxy, since $\lambda_v$ cannot be directly computed and will not have the same meaning to using the $\bar{\delta}_{\rm avg}$ and the $R_{\rm v}$ given by the 2D void finder. Smaller 2D voids are typically void-in-clouds and larger voids tend toward void-in-void, although projection effects make this distinction less precise. Figure \ref{fig:footprints_and_lensing_by_types} illustrates the different types.

To evaluate whether the variation of the Void $\times$ CMB lensing profile across mock families and tracer selections remains consistent when separating by void type, we classify voids into distinct populations and summarize their lensing profiles using three parameters: $\rm{\kappa_{{\rm0}R_v}}$, the lensing signal at the void center; $\rm{\kappa_{\rm max,wall}}$, the peak signal at the compensation wall (when present); and $\rm{\kappa_{\rm \Delta}}$, the net convergence defined as the mean $\kappa$ inside the void minus the mean $\kappa$ in the surrounding region. For 2D voids, this region typically includes the full compensation wall, whereas for 3D voids it may only partially overlap with it. In the plots, the true-scale curve is shown in blue (with values on the $y$-axis), while the others are vertically shifted for visualization purposes. All cases are compared against the mean profile constructed from the \emph{Roman-matched mocks} (see Fig.~\ref{fig:kcharact_rescaled}).

For 3D voids, $\rm{\kappa_{0R_v}}$ increases with $\lambda_v$, reflecting the transition from void-in-void to void-in-cloud types. When $\lambda_v$ is negative (void-in-voids), voids are deeper and produce stronger negative lensing signals. As $\lambda_v$ becomes positive (void-in-clouds), the $\kappa$ profile becomes shallower, the surrounding compensation wall grows stronger, and the net de-magnification $\kappa_{\Delta}$ correspondingly decreases (i.e, becomes more positive).

In 3D voids, $\lambda_v$ provides a physically motivated separation of void-in-void and void-in-cloud populations. While its functional form is calibrated empirically from simulations, the parameter itself traces the underlying gravitational potential \citep{nadathur2017lambda}. In 2D, we rely on $R_{\rm v}$ as a proxy, which is less precise. As a result, binning 2D voids produces a composite profile of objects whose true classifications are ambiguous. We note that this does not undermine the value of 2D voids but for them we still lack a parameter that cleanly separates the underlying populations.

For all 2D void types, $\kappa_{0R_{\rm v}}$ is negative and becomes slightly more negative for larger voids. This follows from the definition of 2D voids, which are identified as negative density contrasts (corresponding to negative convergence values) in the projected galaxy field. The compensation wall signal, $\rm{\kappa_{max,wall}}$, remains nearly constant and also suggests weaker walls compared to highly compensated 3D voids. Finally, $\rm{\kappa_{\Delta}}$ shows that smaller voids, though shallower, exhibit stronger compensation walls and can yield positive values, whereas larger voids, while being deeper, drive $\kappa_{\Delta}$ toward less positive values.

Overall, the tendencies found here agree with \citet{shuster2025_dens}. Voids gradually empty their interiors over time, with matter streaming toward their boundaries. Large, under-compensated voids (void-in-void) in underdense regions evacuate matter efficiently, producing deep cores and weaker net compensation. Smaller, over-compensated voids (void-in-cloud) embedded in high-density environments remain shallower, with matter accumulating near their boundaries, leading to stronger compensation walls and positive $\kappa_\Delta$.

For both 2D and 3D voids, all mock catalogs display nearly identical trends and remain close to the mean profile, reinforcing the idea that the consistency found for the full void sample also holds when separating by type.

\subsubsection{Signal-to-noise ratios}

\begin{figure}[h]
    \centering
    \includegraphics[width=9cm]{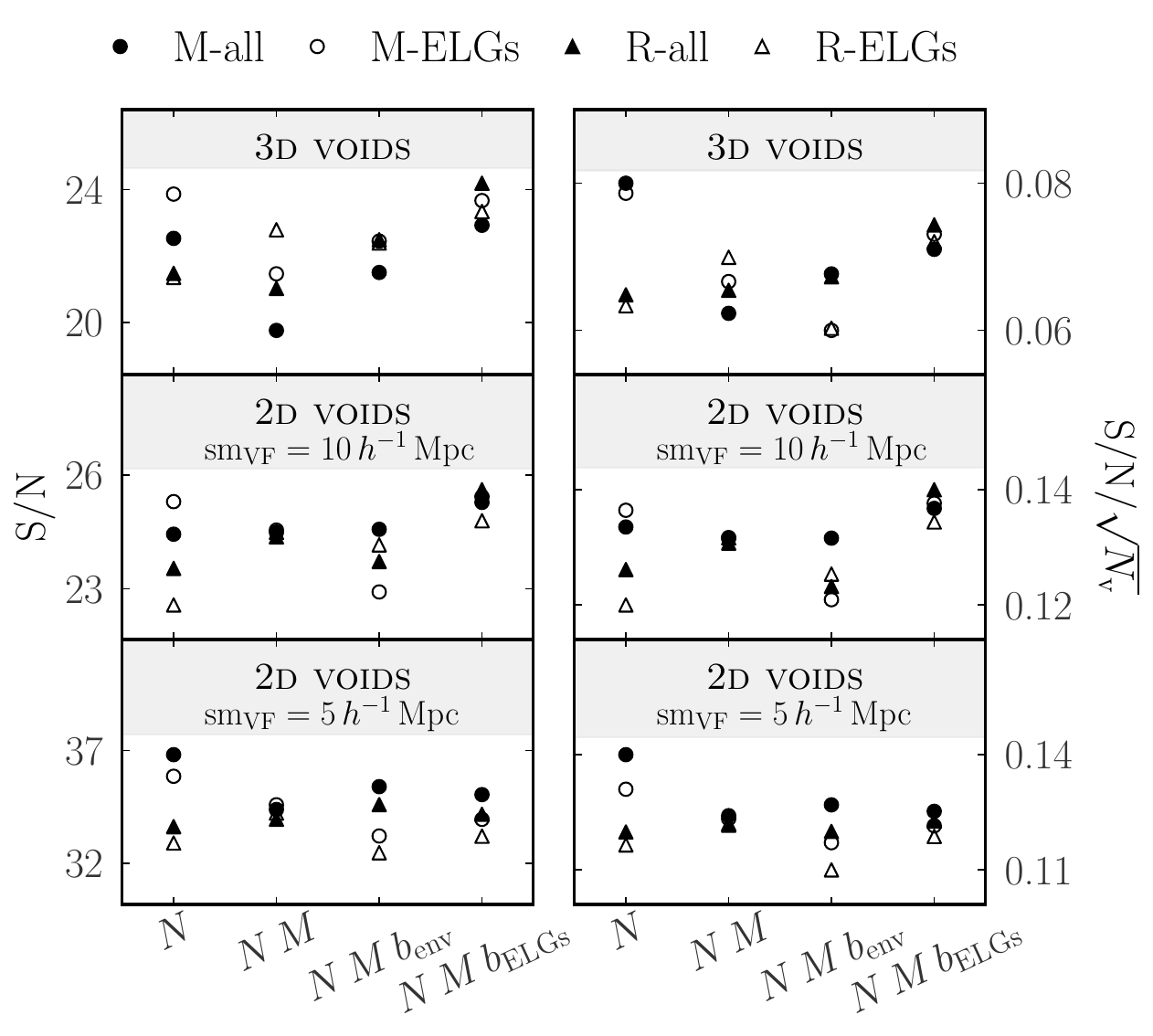} 
    \caption{Comparison of the signal-to-noise ratio (S/N) (left panel) and the signal-to-noise per void (S/N/$\sqrt{N_v}$ ) (right panel) for all mock catalogs and tracer biases.}
    \label{fig:S/N_rescaled}
\end{figure}

Figure~\ref{fig:S/N_rescaled} shows the total S/N (left) and the per-void or normalized S/N (right) across the four mock families and tracer selections. 

For the full set of mocks, both the total and per-void S/N exhibit a characteristic scatter of approximately 10$\%$, which is slightly more pronounced for the normalized S/N in the 3D voids case. Although the lensing profiles in the previous section appear visually more uniform than the S/N distributions, they exhibit variations of a comparable magnitude. The apparent smoothness of the profiles is primarily a result of the statistical uncertainties: the variations lie within the error bars, representing deviations typically below $1\sigma$ and rarely exceeding $2\sigma$.

Restricting to the mocks that match the one- and two-point statistics of the \emph{Roman} reference catalog ($NM$ and $NMb_{\rm ELGs}$), the lensing profiles vary at the level of $\sim1$–$3\%$ for 2D voids and $\sim7$–$9\%$ for 3D voids, while the S/N scatter is reduced to $\sim0.2$–$2\%$ (2D) and below $\sim5\%$ (3D). In contrast, the non-matched families show larger variations, with profile scatter of $\sim6\%$ (2D) and $\sim10$–$15\%$ (3D), and S/N dispersions reaching $\sim10\%$ (2D) and up to $\sim20\%$ (3D).

A main result of this comparison is that 2D voids reach higher S/N values for all mocks catalogs, and once the number of voids is factored out, their efficiency is about twice that of 3D voids. This indicates that projected voids are intrinsically more effective tracers of the CMB lensing signal, as seen in other previous works \citep{vielzeuf2021desy1,camachociurana2024cmbk,demirbozan2024cmbk}. 

The noise properties also show a dependence on the tracer selection, most notably in the simplest mock family, $N$. In this case, we can see how mass-selected samples (M-all, M-ELGs) trace voids more cleanly, resulting in higher S/N per void. By contrast, random-selected samples (R-all,R-ELGs) populate less dense environments which fragments voids and align slightly worse with the lensing signal. For the rest of the cases, the interplay between the mock family and the tracer selection of the input files becomes less straightforward and changes in a non-trivial manner the S/N distribution. 

Ultimately, we note that our goal here is not to identify the catalog with the highest S/N, but to understand in general terms how and why the S/N varies across the different mock catalogs under study. For our analysis, we will adopt the mock that best reproduces the one- and two-point statistics of the \emph{Roman} reference case, regardless of its S/N. If the M-all case shows lower noise because it is a more efficient tracer of CMB lensing, this is valuable for understanding an optimally designed tracer. However, if our real survey lies closer to M-ELGs, then the higher noise observed in this mock is a realistic reflection of what we would expect from the actual survey data.

\subsubsection{Rescale vs Not rescale}

The Void $\times$ CMB lensing signal can be measured using two main approaches: rescaling voids by their radius or stacking them by their angular size \citep{nadathur2016iswstackingtec,kovacs2020iswstackingtec}. Rescaling prioritizes statistical power by normalizing all profiles to a common scale (in units of the void radius), producing uniform stacks where all voids contribute equally to the signal. Although the noise varies across voids due to their different angular sizes and corresponding number of CMB pixels, in this method the lensing signal adds constructively while noise averages down, maximizing the signal-to-noise ratio. 

Non-rescaled stacking, on the other hand, emphasizes physical accuracy, preserving the voids’ intrinsic sizes and environments, though it can align profiles of very different scales and often requires some binning by type or size which reduces its signal-to-noise ratio. 

Regarding the uncertainties of the profiles: inner bins are noisier in both approaches because we are averaging over less pixels, but outer bins behave differently---rescaled profiles show decreasing uncertainties with radius, whereas in non-rescaled profiles uncertainties increase as cosmic variance becomes significant. This holds for the standard radial binning (fixed fraction of $R_v$ or fixed angular width per bin). Alternative schemes (e.g. equal-area or adaptive bins) can redistribute pixel counts and modify this scaling. Finally we note that, since the CMB noise is scale-dependent, non-rescaled stacking maintains a more consistent noise distribution, avoiding the mixing of contributions across scales as in the rescaled method.

\begin{figure}[h]
    \centering
    \includegraphics[width=8.5cm]{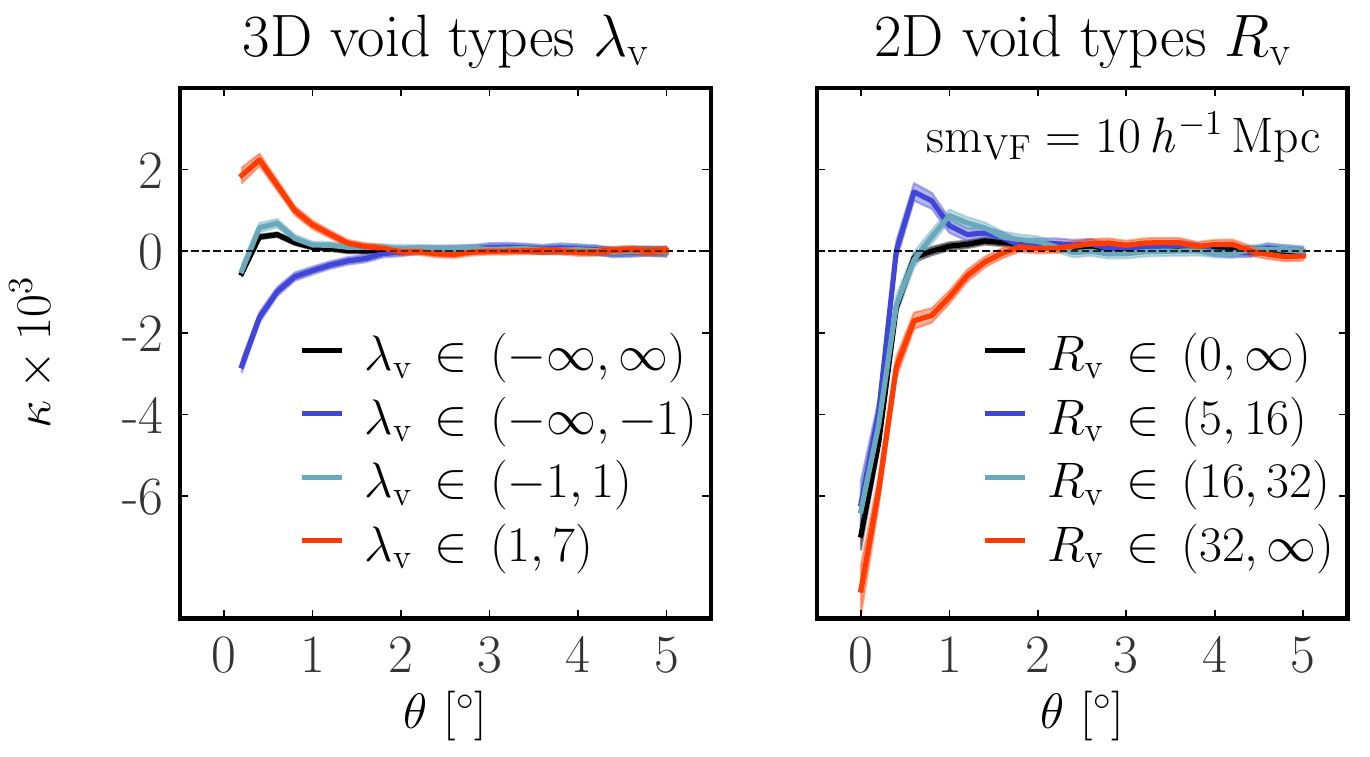} 
    \caption{Non-rescaled Void $\times$ CMB lensing profiles for the mock catalog $N\,M$ (M-all) and various void populations.}
    \label{fig:lensing_by_types_non_rescaled}
\end{figure}

\begin{figure*}[!b]
    \centering
    \includegraphics[width=17cm]{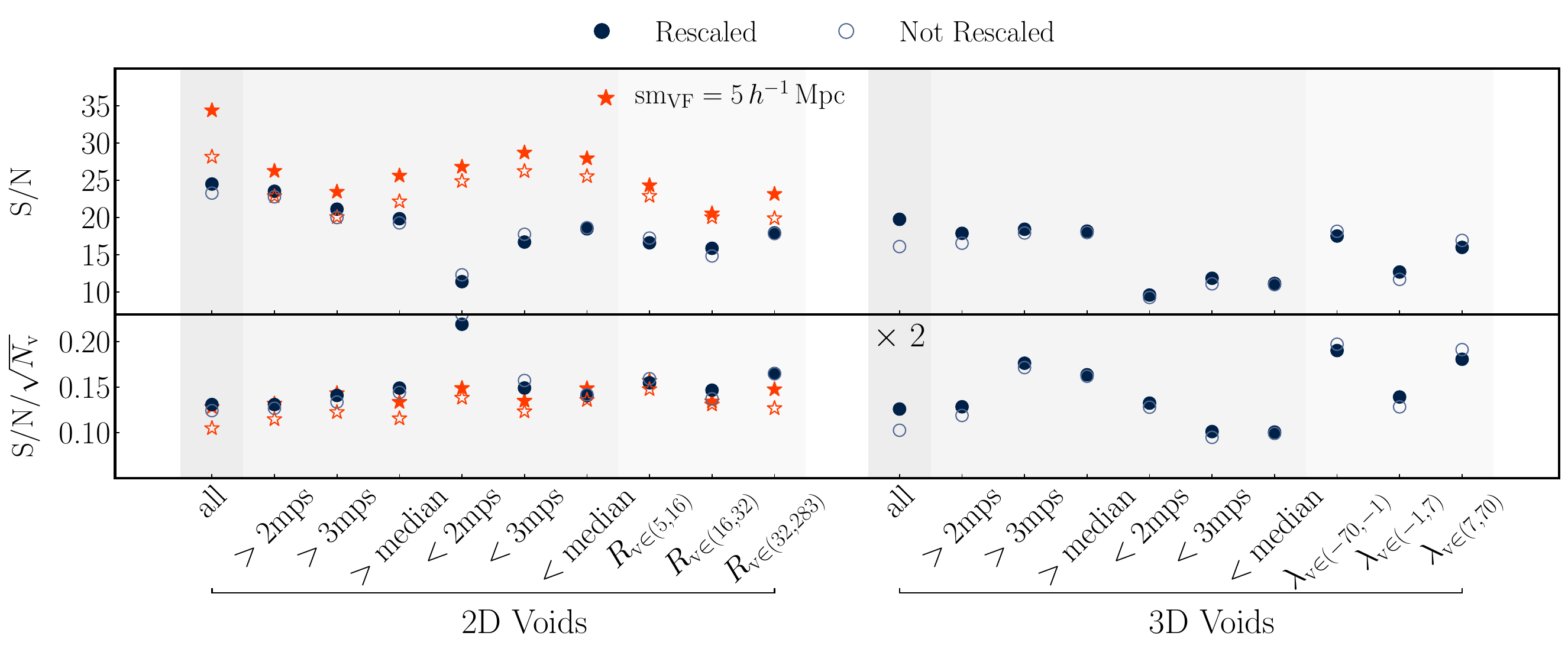}
    \caption{S/N (upper panel) and S/N$\sqrt{N_{\rm v}}$
  (bottom panel) for the mock catalog $N\,M$ (M-all). The results are presented for different radius cuts and void-type divisions. Circles for 2D voids correspond to smoothing scales of ${\rm sm_{VF}} =  10~h^{-1}~$Mpc. In this case the bins for the void types are [11,28], [28,49], and [49,283]~$h^{-1}$~Mpc.}
    \label{fig:S/N_radii_cuts_and_void_types}
\end{figure*}

We repeat the analysis using the non-rescaled methodology (see Appendix~\ref{sec:app_non_rescaled}) and find consistent results. The main conclusions remain unchanged: the strong differences between mock catalogs in one- and two-point statistics are greatly suppressed in Void $\times$ CMB lensing profiles independently of the methodology, underlining the robustness of this observable. 

The main difference appears when separating by void type. For 3D voids, the classification with the parameter $\lambda_v$ provides a clean separation between void-in-void and void-in-cloud types, producing similar profiles for both non-rescale and rescale methodologies. For 2D voids however, where the classification relies on void radius, non-rescaled profiles reveal steep, localized signals for small void-in-clouds and broader, shallower profiles for large void-in-voids. This is a probe of the scaling of the lensing amplitude with a physical mass deficit (which depends on the void size). Rescaling compresses these distinctions, producing an apparently universal profile across void types.

\subsubsection{Mock systematics and cosmological implications}

A $\sim10\%$ variation in the cross-correlation signal across different mock catalogs is not negligible in the context of precision cosmology. While such differences remain statistically indistinguishable under current \emph{Planck} noise levels—typically appearing at less than the $1\sigma$ level—they constitute a potential systematic for next-generation surveys where instrumental noise is significantly reduced. Specifically, a 10$\%$ shift in signal amplitude is comparable to the physical suppression expected from massive neutrinos ($m_{\nu} \approx 0.16$ eV) as seen in \citep{vielzeuf2021desy1}. Our results indicate, however, that the Voids $\times$ CMB lensing signal is less sensitive to specific mock-building nuances than other probes, and if the one- and two-point statistics for both galaxies and voids are consistent with the data (in our case with the reference catalog), the discrepancies in the lensing profiles drop below 10$\%$, which would likely not constitute the cause of any cosmological tension. This difference becomes even more negligible for 2D voids reaching sub percent level with the use of a bigger smoothing scale, opposite of the neutrino effect, where larger smoothing scales typically enhance the physical sensitivity to neutrino mass. 

It is important to note however that in our framework, the one- and two-point statistics for both galaxies and voids emerge as natural byproducts of the mock-generation process rather than being manually tuned to fit observations. If methods such as Halo Occupation Distribution are used to enforce agreement with the observed clustering, part of the cosmological information—such as neutrino effects that affect both the clustering and the lensing signature—may be inadvertently  absorbed into the tracer bias and create artificial tensions. We thus consider that a consistent approach to infer cosmological parameters would be to analyze the clustering and the lensing signal jointly since neutrinos or other cosmological signatures might affect both.

\subsection{Forecast}

We present the Void $\times$ CMB lensing forecast for the \emph{Roman} Space Telescope. We use the $NM$ mock (M-all selection) as our baseline, since is the case that best reproduces the one- and two-point statistics of the \emph{Roman} reference catalog. For this forecast, we first determine the configuration that maximizes the signal-to-noise---comparing the different scaling strategies and void selections---and then use the optimal setup to forecast realistic S/N values under different instrumental noise levels.

\label{sec:results_forecast}
\begin{figure*}[!]
    \centering
    \includegraphics[width=15cm]{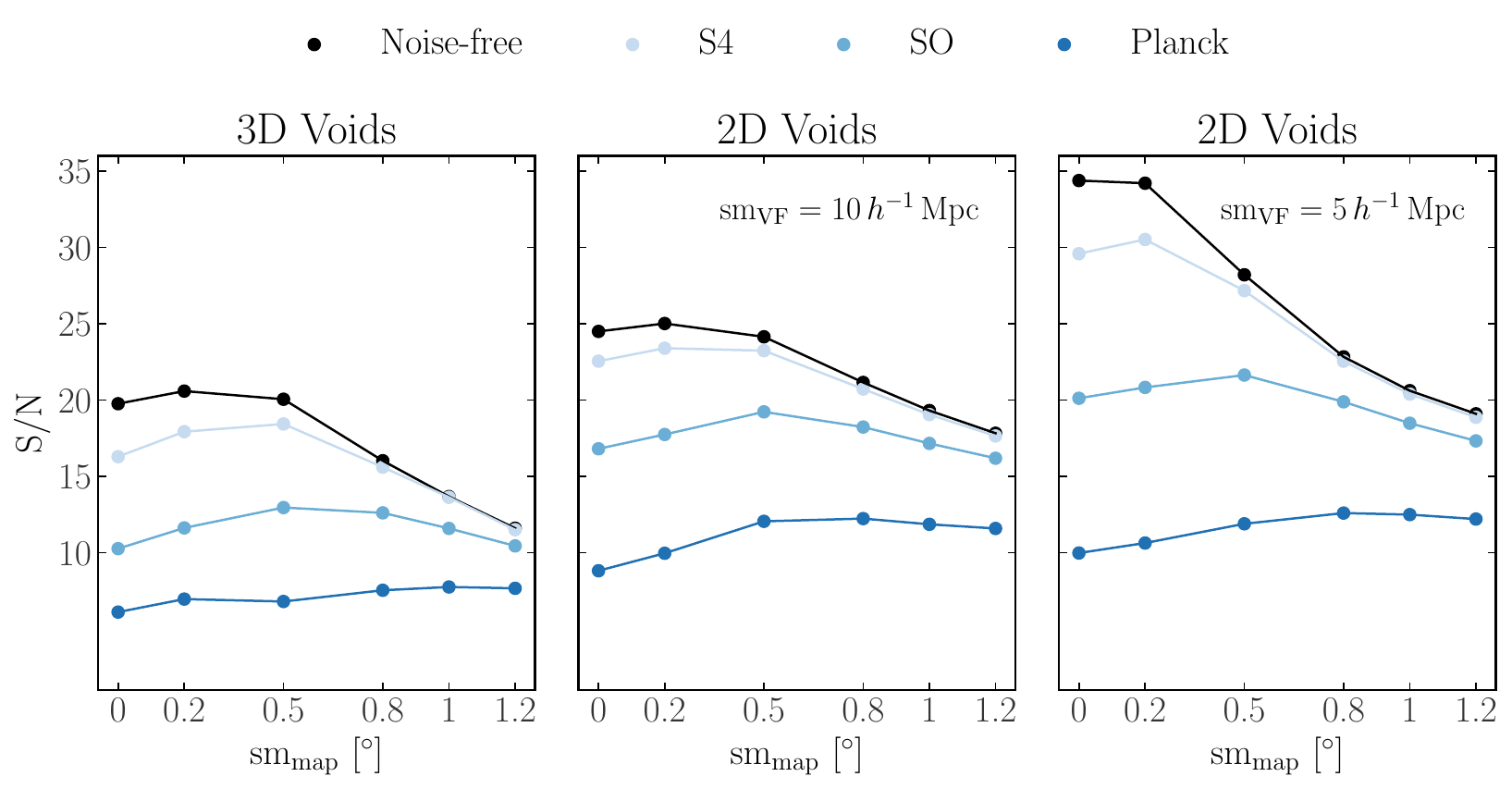} 
    \caption{Forecasted S/N of Void $\times$ CMB lensing for 3D and 2D voids with varying noise levels and CMB smoothing.}
    \label{fig:forecast}
\end{figure*}

\subsubsection{Signal optimization}

Figure~\ref{fig:S/N_radii_cuts_and_void_types} shows the total S/N and the per-void (normalized) S/N/$\sqrt{N_v}$ for rescaled and non-rescaled profiles across different radius cuts and void populations. 

\begin{itemize}
    \item 2D vs 3D voids. The main trend is that 2D voids outperform 3D voids, especially at smaller smoothing scales (${\rm sm_{VF}} = 5~h^{-1}~$Mpc), where the void finder sharpens the contrast of underdensities and increases the number of voids detected. \\
    \item Radius cuts: Regardless of the scaling of the profiles a separate trend appears for 2D and 3D voids when splitting by size due to their different definitions of radius. For 3D voids, the 2 mps (mean particle separation) cut acts as a noise filter that removes spurious detections from shot noise, producing a cleaner sample with higher total S/N and per-void efficiency. For 2D voids, the void radius has two components: the angular radius (influenced by the smoothing scale, ${ \rm sm_{VF}}$) which is what we consider as the void radius, and a fixed line-of-sight radius coming from the slice thickness. The cut does not improve the sample purity but only filters different angular sizes. So for 2D voids, the per-void efficiency remains nearly unchanged across bins, and the total S/N is mainly driven by the number of voids. This is most evident in the smallest bin $<$2mps for ${\rm sm_{VF}}=10~h^{-1}$ Mpc, where the smoothing already exceeds the threshold and the cut removes only $\sim$8$\%$ of the sample. The remaining small voids ($\sim$2700) show an apparently higher per-void efficiency due to their low number compared to the ${\rm sm_{VF}}=5~h^{-1}$ Mpc case ($\sim$32000), despite their comparable lensing depths.\\
    \item Binning vs Full Sample: While subdividing voids by type or size preserves distinct physical profiles and prevents signal dilution, we find that for a Roman-like survey ($\sim$2000 deg$^2$), the loss of statistics per bin generally outweighs the benefits of separating by void type, at least for the rescaled methodology. Consequently, stacking the full void sample maximizes the total S/N for this specific footprint. \footnote{We note that when comparing to real data some works have shown that a joint fit across bins can recover or even increase the cumulative S/N by preserving information that would otherwise be diluted \citep{raghunathan2020cmbk}.} \\
    \item Rescaling vs Non-rescaling: The rescaled methodology provides the highest total S/N when using the full void sample, as it aligns the features of different-sized voids. In contrast, the non-rescaled approach only shows a slight S/N gain when dividing 3D voids by void type. For 2D voids, no improvement is observed when binning because in our main (rescaled) analysis we define bins with equal numbers of voids, which mixes a wide range of radii (e.g., 32--283 $h^{-1}~$Mpc) in a single stack. For the non-rescaled methodology this choice combines intrinsically different profiles and reduces the signal. While previous studies \citep{Raghunathan2019, demirbozan2024cmbk} show that non-rescaled profiles can achieve comparable S/N when combined with CMB filtering techniques (like the matched filter), we focus here on how these methods respond without additional map filtering.
\end{itemize}

\subsubsection{Forecast across instrumental noise levels}

Based on the results above, we adopt the rescaled methodology applied to the full void sample as our baseline for the instrumental noise forecast. Figure \ref{fig:forecast} shows our forecasts results, giving the total signal-to-noise (S/N) for 3D voids and 2D voids across multiple CMB map configurations: the $x$-axis represents the smoothing applied to the CMB map to suppress small-scale instrumental noise, and color markers indicate different noise levels, from none (black) to CMB-S4-like, SO, and {\it Planck}.

\begin{itemize}
    \item Smoothing and Noise Response.  We find that noisier surveys require stronger CMB smoothing to reach peak S/N; for example, \emph{Planck} requires a smoothing scale of $\sim$1$^\circ$, whereas SO and S4-like maps perform optimally at 0.2$^\circ$–0.5$^\circ$. Across all noise regimes, the ${\rm sm_{VF}} = 5~h^{-1}~$Mpc 2D voids remain the best option, though they exhibit higher sensitivity to the choice of mock catalog  (as seen in section \ref{sec:results_impact_mocks_in_lensing_signal}).\\
    
    \item Deviation from noise-free trends. Although we have seen that rescaling the full sample of voids is generally optimal, our comprehensive comparison of stacking techniques and void populations (detailed in Table~\ref{tab:comparison_SNR} of the Appendix) reveals how instrumental noise can alter these trends:
    \begin{itemize}
        \item Rescaling vs Non-rescaling. In the full-sample case, rescaling consistently improves the S/N by $\sim$10–20$\%$ compared to the non-rescaled approach. However, for 3D voids and \emph{Planck} noise, the non-rescaled method yields a 33$\%$ higher S/N. This is likely because rescaling upweights the high-multipole noise associated with smaller, more numerous voids—an effect to which the 3D sample is particularly sensitive.\\
        \item Binning vs Full Sample: In the noise-free case, stacking the full sample is preferred. However, as instrumental noise increases, binning 3D voids by type becomes increasingly effective for both rescaled and non-rescaled methodologies, specially for noise levels as SO and \emph{Planck}. This suggests that grouping similar profiles provides a variance reduction that outweighs the loss of statistical power, particularly for the deepest underdensities. For 2D voids, dividing by type still shows no S/N improvement in either methodology and noise scheme. For the rescaled methodology we do not have a clean separation of void types that boost the signal and for the non-rescaled methodology the binning is too wide to produce distinct profiles.
    \end{itemize}
\end{itemize}

In the future, matched filter techniques could potentially enhance the S/N of binned voids in the non-rescaled method to match the rescaled approach \citep{raghunathan2020cmbk,demirbozan2024cmbk}. This would allow a clearer separation of void types without sacrificing the precision. Moreover, these filtering techniques reduce the map noise by weighting scales according to signal-to-noise, rather than manually applying Gaussian smoothing to suppress small-scale fluctuations, as we do for the rescaled methodology. For \emph{Planck}-level noise their improvement over standard stacking is modest, but for future low-noise surveys such as CMB-S4-like they are expected to become essential for fully exploiting the available statistical power. In this work we present a conservative forecast based on the rescaled methodology, for which matched filtering is not directly applicable, and defer optimized analyses to future studies including realistic noise realizations.

\section{Conclusions}
\label{sec:conclusions}
The path toward establishing Void $\times$ CMB lensing as a precision cosmological probe requires both an understanding of its dependence on cosmological parameters and of the impact of methodological choices on its measurement. In this work, we have focused on the latter, specifically tailored for the high-density regime of the \emph{Roman} Space Telescope. \\

To this end, in a companion paper \citep{perezsar2025_roman_mocks} we develop and validate a method to replicate mock catalogs from a given parent simulation for other simulations of interest such as those with associated CMB maps. There, we systematically analyze how the different catalogs of varying complexity affect void and galaxy statistics. The resulting mocks constitute the datasets of this study, where we examine their impact on the Void $\times$ CMB lensing signal and its associated noise properties. In this work, we compare 2D and 3D void-finding algorithms and measurement strategies, and produce forecasts for the \emph{Roman} Telescope in combination with current and next-generation CMB experiments ({\it Planck}, SO, CMB-S4-like).\\

We note that while this study utilizes a single sky realization, this choice provides a controlled environment to isolate methodological systematics. With the underlying large-scale structure (LSS) fixed, we ensure that the observed variations in the lensing signal are driven strictly by mock-generation techniques rather than by stochastic fluctuations. Cosmic variance can still affect the absolute amplitude, but the relative differences we find reflect the effect of the modeling choices alone.\\

Our conclusions are:

\paragraph{\textbf{Stability of the Void $\times$ CMB lensing signal to mock catalog features.}} The Void $\times$ CMB lensing signal itself remains stable across different mock catalogs, even when galaxy and void clustering properties vary substantially. Since CMB lensing probes the underlying total mass distribution, the signal is less sensitive to the fine details of galaxy assignment or galaxy bias. We find that the dispersion in the profiles across the full suite of mocks is typically of the order of $\sim 10\%$. At the noise levels of current experiments like \emph{Planck}, or even SO, these variations are statistically sub-dominant and unlikely to pose any tensions. However, for next-generation experiments (e.g., CMB S4-like), a 10$\%$ shift becomes  significant, as it is comparable to some cosmological signature such as the suppression in the signal expected from massive neutrinos ($m_{\nu} \approx 0.16$ eV) \cite{vielzeuf2023_probe_n}. 

Crucially, we find that the differences between mocks are significantly diminished for 2D voids. When restricting our analysis to mocks that correctly reproduce the one- and two-point statistics of the \emph{Roman} reference catalog (the $NM$ and $NMb_{\rm ELGs}$ families), the lensing profile variations drop to $\sim 1$--$3\%$ for 2D voids and the S/N scatter is reduced to $\sim 0.2$--$2\%$. Furthermore, we observe that as the smoothing scale increases, the sensitivity to mock-building nuances decreases, whereas the physical sensitivity to the neutrino mass is reported to increase. This divergent behavior suggests that 2D void-lensing is particularly well-suited for breaking the degeneracy between methodological systematics and new physics.

Rescaling and non-rescaling methodologies perform similarly, though non-rescaling generally yields a lower signal-to-noise ratio. The use of matched filters could potentially mitigate this difference, but lies beyond the scope of this work. We also find that while 2D void finders provide higher signal-to-noise (about twice when factoring out the void abundance), 3D voids offer a clearer distinction between different void environments. This is valuable for testing the reliability of mocks by confronting the environment-dependent predictions with real observations and it is also useful for physical interpretation and theory development.

Beyond maximizing the S/N while reducing catalog-specific effects, it is important to identify which voids carry the strongest discriminatory power—such as the 2D voids used for modified gravity tests in \cite{paillas2019} or for studying the neutrino effect \cite{vielzeuf2021desy1}. A systematic study of how the Void $\times$ CMB lensing signal responds to a broader parameter space under different stacking strategies and void definitions is a necessary step for establishing voids as a precision probe.

\paragraph{\textbf{Forecasts for the \emph{Roman} telescope using the best mock catalog case.}} Our forecasts show that current {\it Planck} data are noise-limited and do not provide competitive constraints for the \emph{Roman} Space Telescope (see below). However, upcoming CMB experiments such as the Simons Observatory (SO) and CMB-S4-like surveys will deliver substantial improvements in sensitivity. Our main results are summarized in Table \ref{tab:sum_results_SNR}.

\begin{table}[h!]
\hspace*{-0.5cm} 
\resizebox{0.50\textwidth}{!}{
\begin{tabular}{lcclclcl}
\multicolumn{1}{c}{} &  & \multicolumn{6}{c} {\textbf{Noise configuration}} \\  \\ \cline{3-8} 
\multicolumn{1}{c}{} & \multicolumn{1}{c|}{} & \multicolumn{2}{c|}{\textbf{S4-like}} & \multicolumn{2}{c|}{\textbf{SO}} & \multicolumn{2}{c|}{\textbf{\emph{Planck}}} \\ \cline{3-8} 
\multicolumn{1}{c}{} & \multicolumn{1}{c|}{} & S/N & \multicolumn{1}{c|}{sm$_{\rm map}$ {[}º{]}} & S/N & \multicolumn{1}{c|}{sm$_{\rm map}$ {[}º{]}} & S/N & \multicolumn{1}{c|}{sm$_{\rm map}$ {(}º{)}} \\ \cline{2-8} 
\multicolumn{1}{l|}{\multirow{7}{*}{\rotatebox{90}{\textbf{Void type}}}} & \multicolumn{1}{l|}{} & \multicolumn{1}{l}{} & \multicolumn{1}{l|}{} & \multicolumn{1}{l}{} & \multicolumn{1}{l|}{} & \multicolumn{1}{l}{} & \multicolumn{1}{l|}{} \\ 
\multicolumn{1}{l|}{} & \multicolumn{1}{c|}{\textbf{3D voids}} & 18 & \multicolumn{1}{c|}{0.5} & 13 & \multicolumn{1}{c|}{0.5} & 8 & \multicolumn{1}{c|}{1} 
\\
\multicolumn{1}{l|}{} & \multicolumn{1}{c|}{\begin{tabular}[c]{@{}c@{}}\textbf{2D Voids}\\ \tiny{(sm$_{\rm VF}$ = 10 $h^{-1}$ Mpc)}\end{tabular}} & 23 & \multicolumn{1}{c|}{0.2} & 19 & \multicolumn{1}{c|}{0.5} & 12 & \multicolumn{1}{c|}{0.8} 
\\
\multicolumn{1}{l|}{} & \multicolumn{1}{c|}{\begin{tabular}[c]{@{}c@{}}\textbf{2D Voids}\\ \tiny{(sm$_{\rm VF}$ = 5 $h^{-1}$ Mpc)}\end{tabular}} & 31 & \multicolumn{1}{c|}{0.2} & 22 & \multicolumn{1}{c|}{0.5} & 13 & \multicolumn{1}{c|}{0.8} \\
\multicolumn{1}{l|}{} & \multicolumn{1}{l|}{} & \multicolumn{1}{l}{} & \multicolumn{1}{l|}{} & \multicolumn{1}{l}{} & \multicolumn{1}{l|}{} & \multicolumn{1}{l}{} & \multicolumn{1}{l|}{} \\ \cline{2-8} 
\end{tabular}}
\caption{Maximum S/N and associated CMB smoothing scale for the \emph{Roman} Voids $\times$ CMB lensing cross-correlations. Results span across different noise configurations and void finders using the full 2D and 3D void samples without dividing into populations.}
\label{tab:sum_results_SNR}
\end{table}

The optimal smoothing scale is survey- and void finder-dependent when applying the rescaled methodology. Compared to results from other works \citep[e.g.,][who combines DESI DR9 3D voids with {\it Planck}]{sartori2024cmbk}  that already report $\rm \sim 17\sigma$, our findings indicate that this observable benefits more from larger survey footprints than from extreme depth over limited areas.

Although \emph{Roman}’s relatively smaller sky coverage may limit standalone lensing constraints, its depth, tracer density, and angular resolution will enable the detection of smaller voids and improved mapping of their internal structure. This extends void studies beyond the capabilities of previous wide but relatively sparse surveys. As analyses transition to denser, high-resolution data, methodological aspects become increasingly important to improve both the detection significance and the scientific return. Our study contributes to this broader effort by testing the best methodological configuration for \emph{Roman}'s Voids $\times$ CMB lensing measurements. Equally important is the planned synergy of \emph{Roman} with other surveys and cosmological probes, critical for obtaining robust constraints and mitigating systematics. Combined with forthcoming CMB datasets, as well as with large-scale structure measurements from e.g. DESI, Euclid and LSST, \emph{Roman}’s observations will deliver complementary and high-fidelity constraints on cosmic structure and evolution \citep{eifler2020,wenzl2022,eifler2024}. In this exciting scenario, Void $\times$ CMB lensing has the potential to place direct constraints on cosmological parameters for the first time.

\section{Acknowledgements}
This paper made use of the IAC HTCondor facility
(\url{http://research.cs.wisc.edu/htcondor/}), partly financed by the Ministry of Economy and Competitiveness with FEDER funds, code IACA13-3E-2493. MP wishes to acknowledge the contribution of the IAC High-Performance Computing support team and hardware facilities to the results of this research, specially to Ángel de Vicente. 

MP acknowledges Giulio Fabbian for useful discussions during the early stages of this work, as well as support from the pre-doctoral program at the Center for Computational Astrophysics, Flatiron Institute. Research at the Flatiron Institute is supported by the Simons Foundation. MP acknowledges support from the Agencia Estatal de Investigación del Ministerio de Ciencia en Innovación (AEIMICIN) and the European Social Fund (ESF+) under grant PRE2021-098156. MP and C.H.-M acknowledge the support of the Spanish Ministry of Science and Innovation under the grants PID2021-126616NB-I00 and "DarkMaps" PID2022-142142NB-I00, and from the European Union through the grant "UNDARK" of the Widening participation and spreading excellence program (project number 101159929).

The Large-Scale Structure (LSS) research group at Konkoly Observatory has been supported by a \emph{Lend\"ulet} excellence grant by the Hungarian Academy of Sciences (MTA). This project has received funding from the European Union’s Horizon Europe research and innovation programme under the Marie Skłodowska-Curie grant agreement number 101130774. Funding for this project was also available in part through the Hungarian National Research, Development and Innovation Office (NKFIH, grant OTKA NN147550). 

AP acknowledges support from the European Research Council (ERC) under the European Union’s Horizon programme (COSMOBEST ERC funded project, grant agreement 101078174), as well as support from the French government under the France 2030 investment plan, as part of the Initiative d’Excellence d’Aix-Marseille Université - A*MIDEX AMX-22-CEI-03 as well as support from the Center for Computational Astrophysics, the Flatiron Institute and the Simons Foundation for early stages of this work.
YW gratefully acknowledges support from NASA Grant 80NSSC24M0021, "Project Infrastructure for the Roman Galaxy Redshift Survey".

\bibliography{bibliography}{}
\bibliographystyle{aasjournal}

\appendix

\section{Comparison of void catalogs}
\label{sec:app_void_catalogs}

\begin{figure}[h!]
    \centering
    \hspace*{-1cm}
    \includegraphics[width=7.5cm]{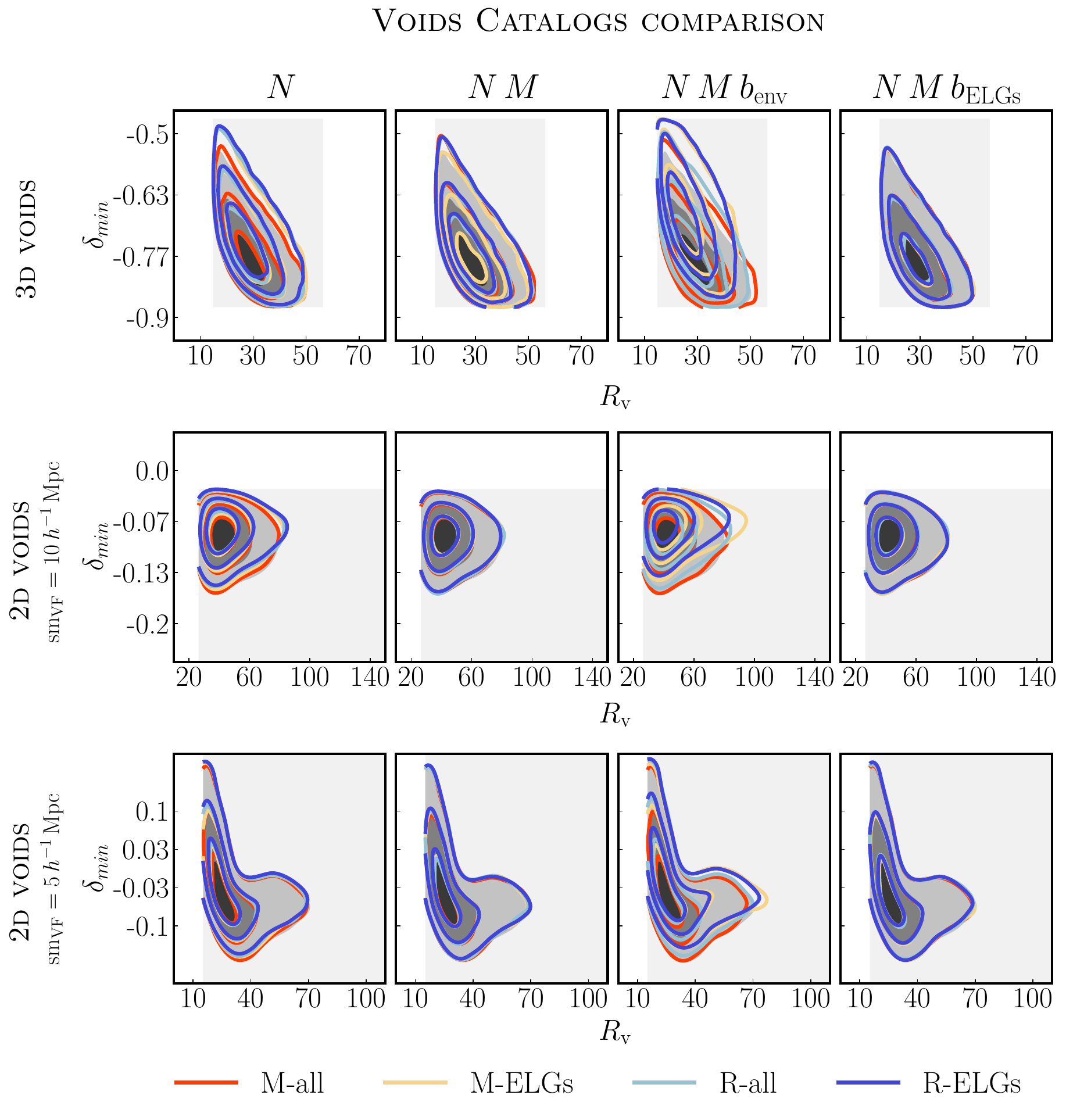}
    \caption{Distribution of void minimum density and radius for the different catalogs. Columns correspond to mock families, while colors show different tracer selections.}
    \label{fig:comparison_void_catalogs}
\end{figure}

We summarize the main properties of the void catalogs introduced in \cite{perezsar2025_roman_mocks} through their minimum density and effective radius. Figure~\ref{fig:comparison_void_catalogs} shows contour distributions for each mock family (columns) and tracer selection (colors), highlighting the level of agreement of the catalogs in comparison with the reference \emph{Roman} mock from \cite{zhai2021clustering}, depicted with grey-shaded contours.

\begin{figure*}[!b]
    \centering
    \includegraphics[width=11.5cm]{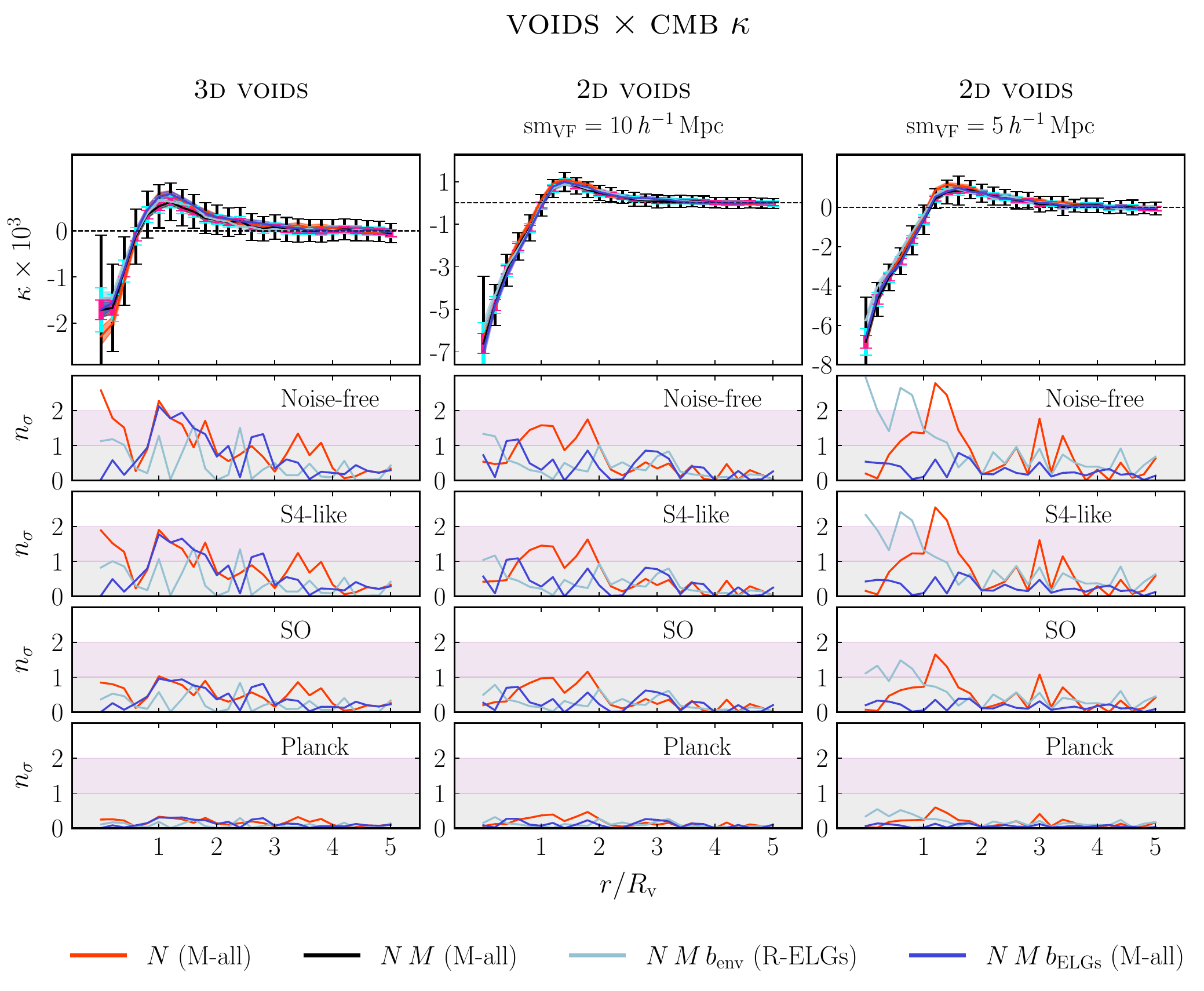}
    \caption{Void $\times$ CMB $\kappa$ signal under different instrumental noise levels (\emph{Planck}, SO, CMB-S4-like). Top: stacked profiles. Bottom: residuals and corresponding tension.}
    \label{fig:signalk_with_different_IN}
\end{figure*}

\section{Void $\times$ CMB $\kappa$ with instrumental noise}
\label{sec:app_signal_with_IN}
Figure~\ref{fig:signalk_with_different_IN} shows the Void $\times$ CMB $\kappa$ signal for the whole void sample including instrumental noise. We consider the three void definitions (3D voids, and 2D voids with ${\rm sm_{VF}} = 10$ and $5~h^{-1}~{\rm Mpc}$) and a set of mock catalogs that cover both well-matched and discrepant cases, based on their ability to reproduce the void and galaxy one- and two-point statistics of the reference \emph{Roman} mock from \cite{zhai2021clustering} (more details in \citealt{perezsar2025_roman_mocks}). Deviations at the $\sim2\sigma$ level appear only for S4-like noise, and marginally for SO in the 2D catalog with ${\rm sm_{VF}} = 5~h^{-1}~{\rm Mpc}$. These are driven by mocks that fail to reproduce the reference statistics (e.g., $N$ (M-all) and $N\,M\,b_{\rm env}$ (R-ELGs)) and would therefore be excluded from a Void $\times$ CMB $\kappa$ analysis. Once these statistics are matched, differences between catalogs remain below $1\sigma$ and are unlikely to generate any significant tension with $\Lambda$CDM.

\section{Void $\times$ CMB $\kappa$ with several radius cuts}
\label{sec:app_signal_with_radius_cuts}
\begin{figure*}[!t]
    \centering
    \includegraphics[width=19cm]{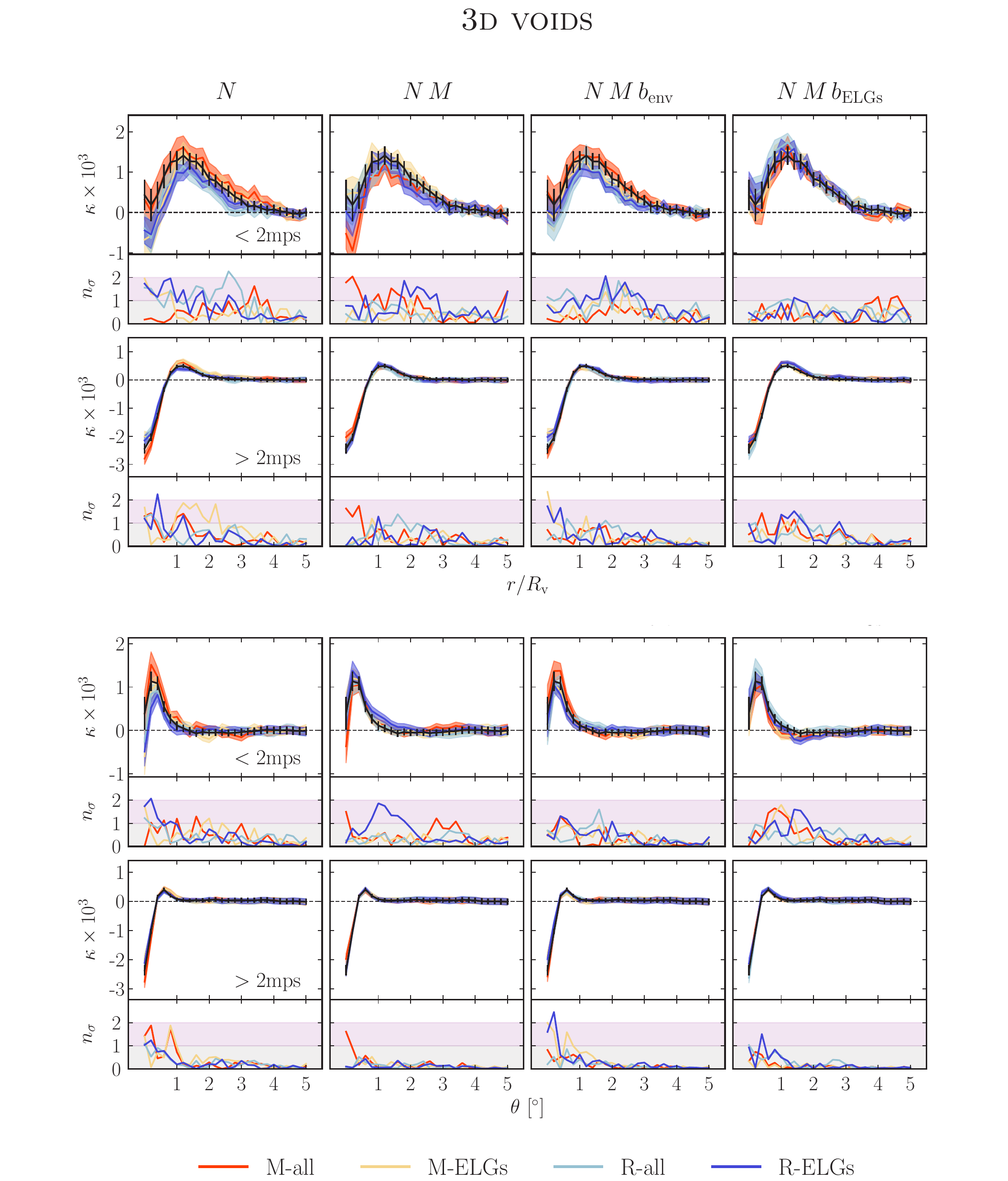}
    \caption{Rescaled (top) and non-rescaled (bottom) lensing profiles for 3D voids under radius cuts above/below twice the mean particle separation (mps). The residuals indicate the difference of each profile to the mean profile resulting from the cases that matched the one-point and two-point statistics of the reference catalog.}
    \label{fig:lensing_rescaled_and_not_rescaled_with_Rcuts_3d}
\end{figure*}

\begin{figure*}[!h]
    \centering
    \includegraphics[width=19cm]{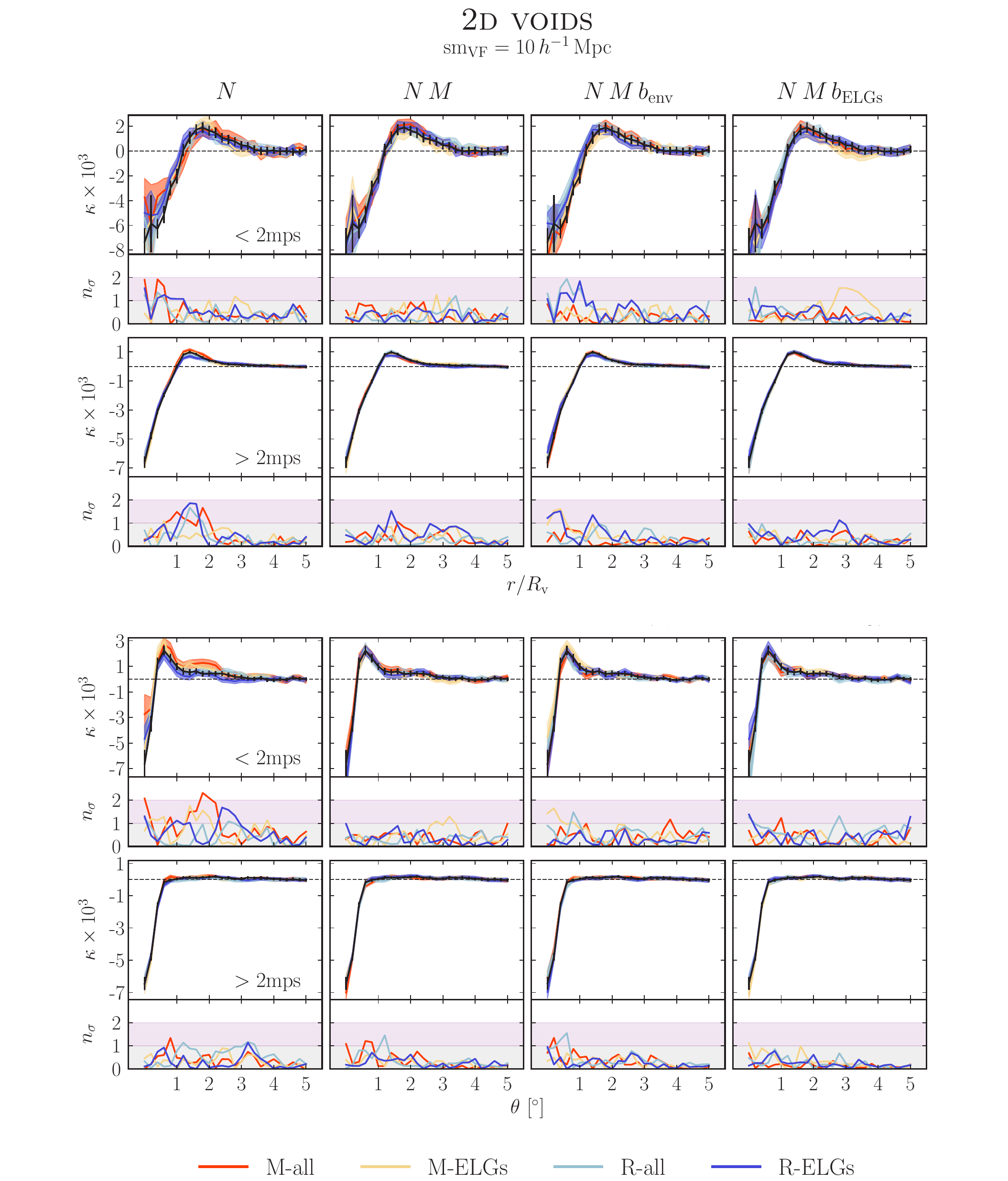}
    \caption{Same as Figure \ref{fig:lensing_rescaled_and_not_rescaled_with_Rcuts_3d} for 2D voids ($\rm sm_{VF} = 10~h^{-1}~\mathrm{Mpc}$).}
    \label{fig:lensing_rescaled_and_not_rescaled_with_Rcuts_2dsm10}
\end{figure*}

\begin{figure*}[!h]
    \centering
    \includegraphics[width=19cm]{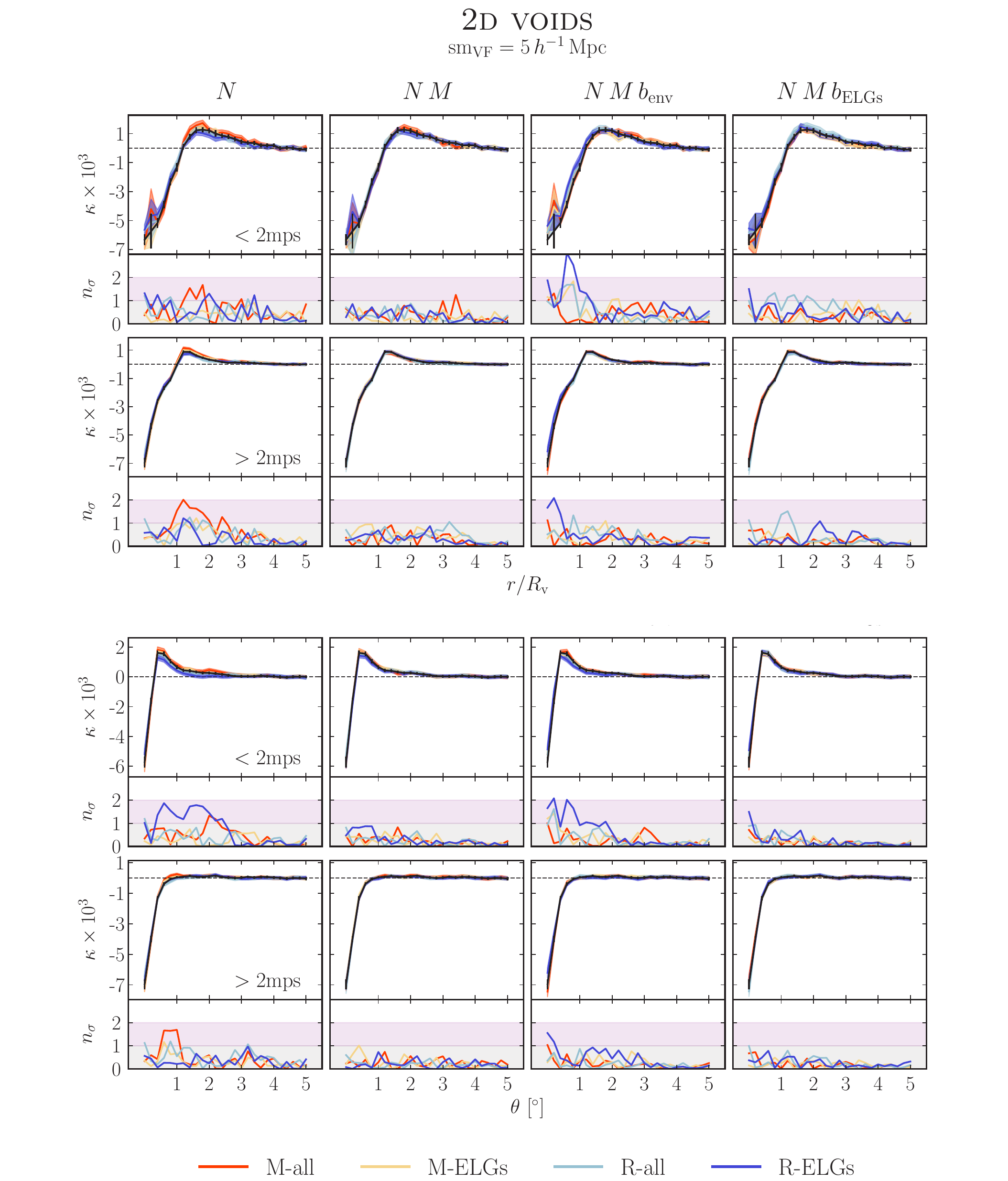} 
    \caption{Same as Figure \ref{fig:lensing_rescaled_and_not_rescaled_with_Rcuts_3d} for 2D voids ($\rm sm_{VF} = 5~h^{-1}~\mathrm{Mpc}$).}
    \label{fig:lensing_rescaled_and_not_rescaled_with_Rcuts_2dsm5}
\end{figure*}

We explore the impact of void size cuts in the signal by splitting the sample above and below twice the mean particle separation (mps) for both rescaled and non rescaled profiles (see \Cref{fig:lensing_rescaled_and_not_rescaled_with_Rcuts_3d,fig:lensing_rescaled_and_not_rescaled_with_Rcuts_2dsm10,fig:lensing_rescaled_and_not_rescaled_with_Rcuts_2dsm5}).

Voids smaller than 2 mps show larger scatter for two independent reasons. First, they contain fewer tracers, which increases the shot noise and the fraction of spurious detections in the void identification. Second, their small angular extent provides fewer pixels to average per radial bin, increasing the statistical uncertainty in the stacked profiles. Despite their intrinsic scatter, small voids do not appear to drive the catalog-to-catalog variations since differences are also present in the inner bins of larger voids. The discrepancies therefore appear to be driven more generally by differences in the tracer bias among the mocks.

\begin{figure*}[!h]
    \centering
    \includegraphics[width=13cm]{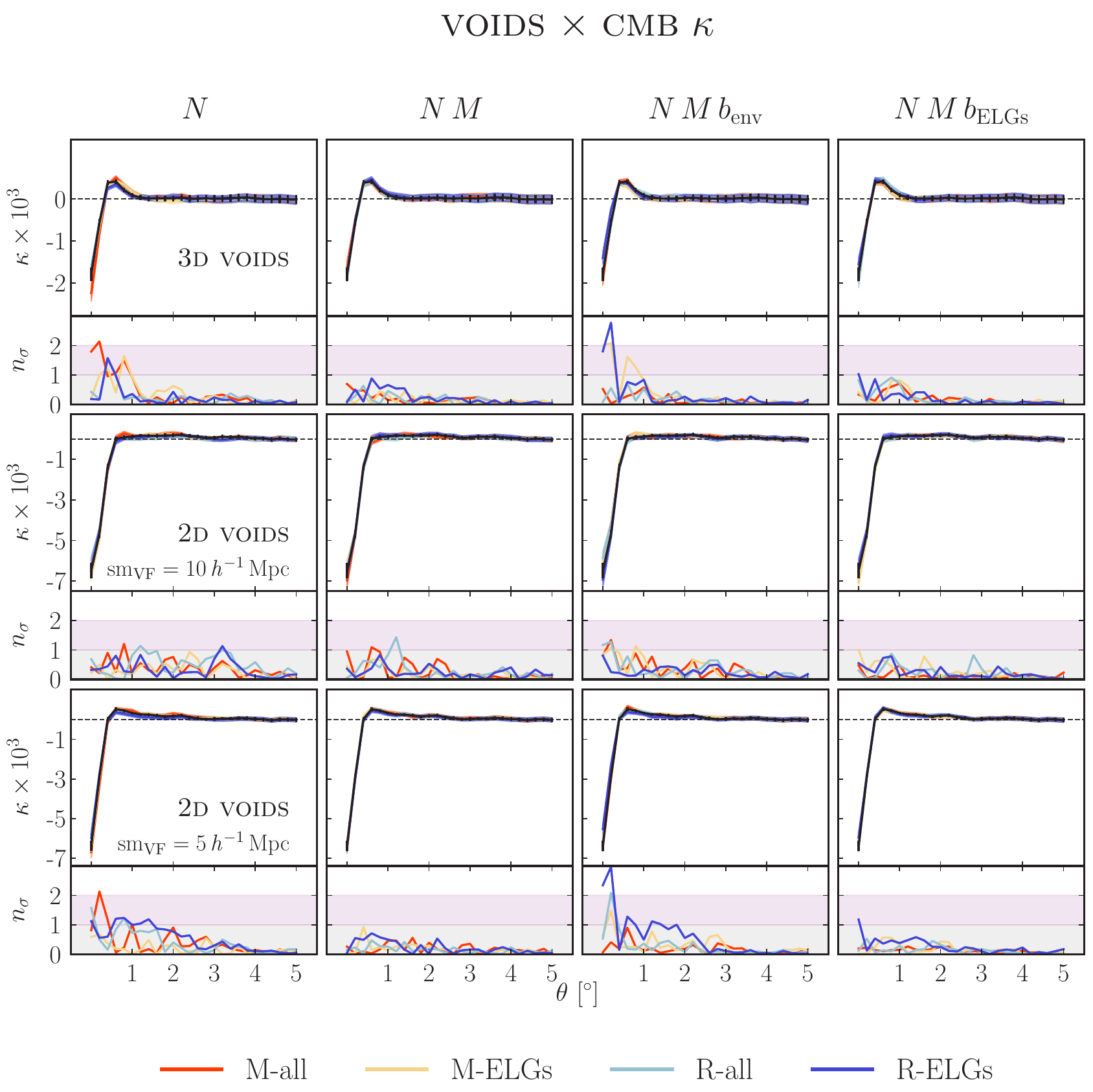}
    \caption{Same as Figure \ref{fig:lensing_comparison_main} but non-rescaled.}
    \label{fig:lensing_comparison_nonrescaled}
\end{figure*}

{\color{white}{.}}\\\\\\
\section{Void $\times$ CMB $\kappa$ signal for the non-rescaled methodology}
\label{sec:app_non_rescaled}

We show the equivalent study performed in Sect.~\ref{sec:results_impact_mocks_in_lensing_signal} for the non-rescaled methodology, where void profiles are measured as a function of angular separation from the void center rather than scaled by radius. From \Cref{fig:lensing_comparison_nonrescaled,fig:S/N_nonrescaled,fig:kcharact_nonrescaled} we show the Void $\times$ CMB $\kappa$ signal when considering the whole void sample (covering 3D voids and 2D voids identified with ${\rm sm_{VF}} = 10~h^{-1}$ Mpc and ${\rm sm_{VF}} = 5~h^{-1}$ Mpc); the same but dividing by void type; and the signal-to-noise ratios of the profiles for the different mock families and tracer selections.

\begin{itemize}
\item If we compare Fig.~\ref{fig:lensing_comparison_nonrescaled} with Fig.~\ref{fig:lensing_comparison_main}, 3D voids tend to exhibit higher average noise than non-rescaled ones. One might think that this is likely to result from the mixing of CMB noise across scales. However, in the non-rescaled case, profiles at larger radii (above 2–3$^\circ$) average over relatively large scales that are common for all cases, washing out differences between mock catalogs and making the profiles appear artificially similar. When restricted to 3$^\circ$, the noise levels of the two methods are broadly comparable.\\

\begin{figure}[h]
    \centering
    \includegraphics[width=8cm]{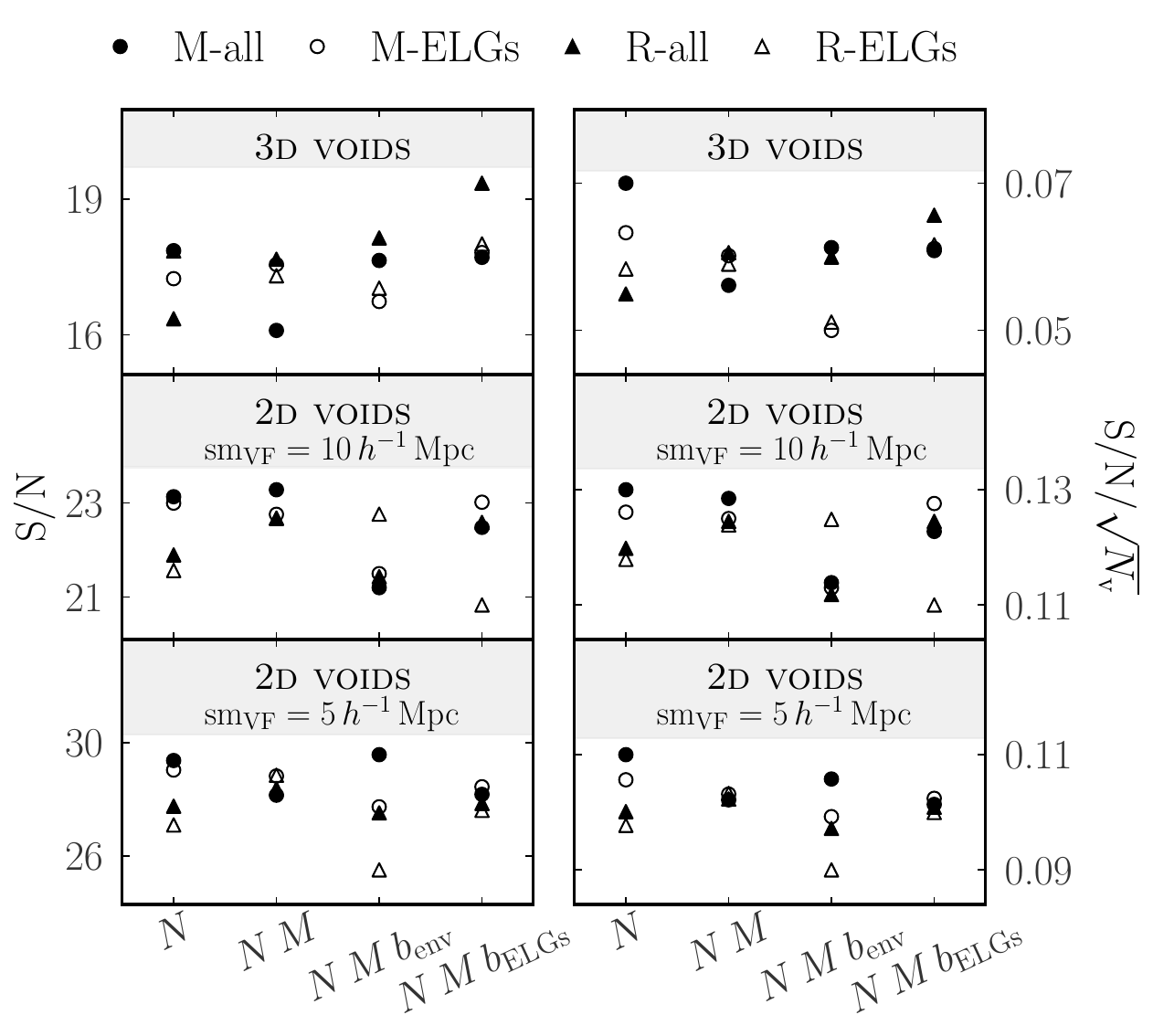} 
    \caption{Same as Figure \ref{fig:S/N_rescaled} but non-rescaled.}
    \label{fig:S/N_nonrescaled}
\end{figure}

\begin{figure*}[h]
    \centering
    \hspace*{-0.6cm} 
    \includegraphics[width=19.5cm]{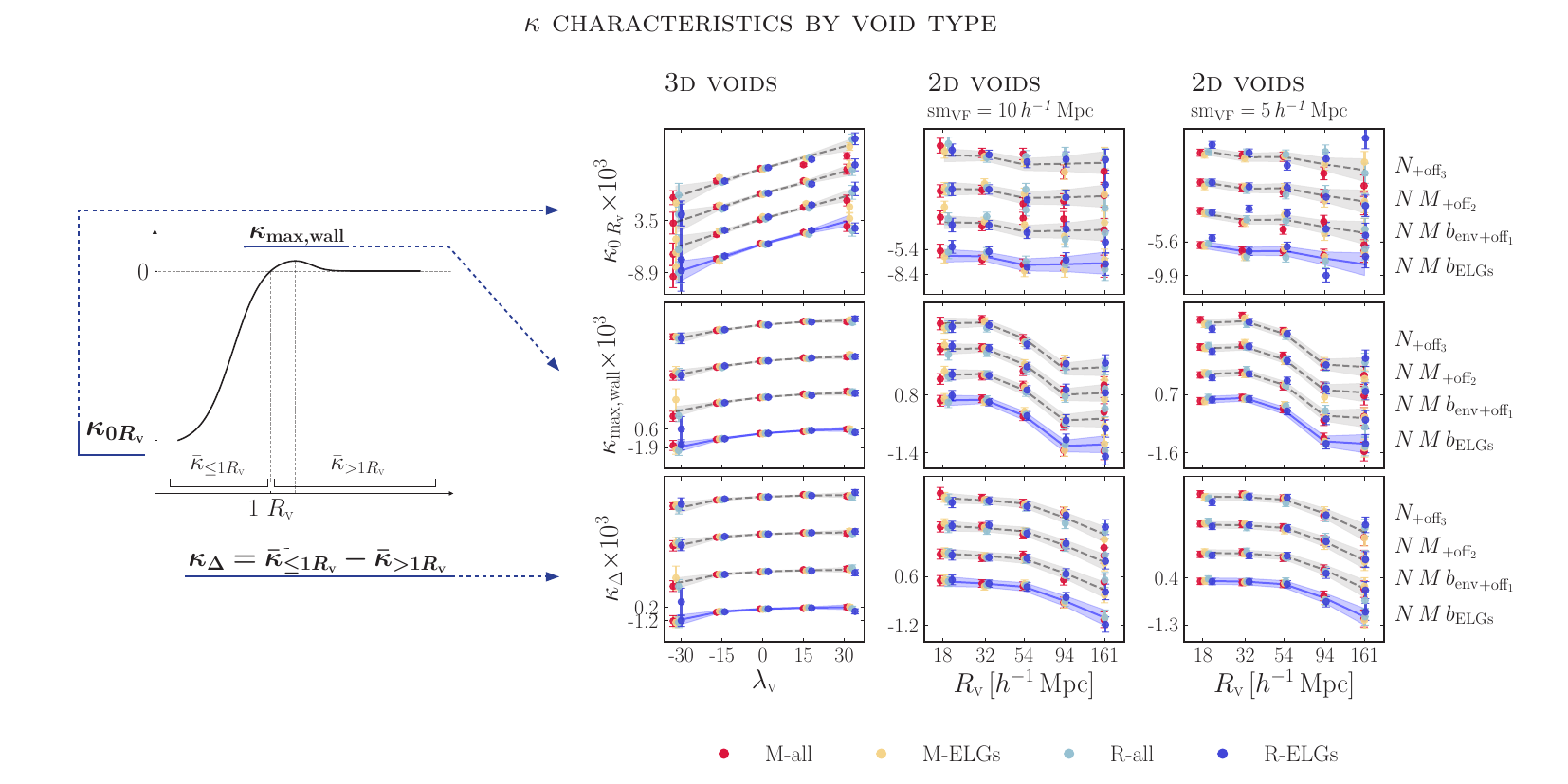}
    \caption{Same as Figure \ref{fig:kcharact_rescaled} but non-rescaled.}
    \label{fig:kcharact_nonrescaled}
\end{figure*}

\item The signal-to-noise trends across the different mock families and tracer selections remain consistent with the rescaled analysis and are reported in Figure \ref{fig:S/N_nonrescaled}.\\

\item If we separate by void type in \ref{fig:kcharact_nonrescaled} the main difference with respect the rescaled analysis is that for 2D voids the distinction of void populations is clearer. This impacts the values of the $\kappa_{\rm max,wall}$ (which was almost constant and without a definite trend in the rescaled case) and $\kappa_{\rm \Delta}$, which now show an evolution. As the void population shifts towards larger and more under compensated voids (voids-in-voids), the compensation wall amplitude and the net convergence become progressively more negative. \\

\end{itemize}

\section{Forecast table}
\label{sec:app_forecast_table}

\Cref{tab:comparison_SNR} summarizes the signal-to-noise ratios of \emph{Roman} Void × CMB $\kappa$ measurements across different noise configurations (\emph{Planck}, SO, and CMB-S4–like surveys), void definitions (2D and 3D), stacking methods (rescaled and non-rescaled profiles), and CMB map smoothing scales. \\

When considering all voids, the non-rescaled approach generally produces a lower S/N compared to the rescaled method, which is consistent with our previous findings: stacking non-rescaled voids mixes very different void profiles, diluting the overall signal. On average, the rescaled method achieves a 21~$\%$ higher S/N for 3D voids (except for 3D voids in {\it Planck}, where non-rescaled is 33~$\%$ higher \footnote{This is likely because rescaling upweights the high-multipole (small-scale) noise associated with the smaller, more numerous voids and 3D voids, due to their lower abundance and weaker signal, are more sensitivity to this effect.}), 16~$\%$ higher for 2D voids with ${\rm sm_{VF}} =5~h^{-1}~$Mpc, and 12~$\%$ higher for 2D voids with ${\rm sm_{VF}} =10~h^{-1}~$Mpc. When binning by void type, the differences in S/N between rescaled and non-rescaled methods are much smaller and more variable: for 3D voids, S/N ranges roughly from –9 to 5~$\%$, while for 2D voids it spans approximately –2 to 11~$\%$, depending on void size and smoothing scale.\\

When voids are separated by type, the rescaled method yields a slightly lower total S/N than stacking the full sample, except for the most extreme 3D bins. For these bins, {\it Planck} shows gains of 49~$\%$ ($\lambda_v<-1$) and 27~$\%$ ($\lambda_v>7$), and SO gains 23~$\%$ ($\lambda_v<-1$) and 4~$\%$ ($\lambda_v>7$) relative to the full void sample. In the non-rescaled method, 3D voids consistently gain S/N across all noise levels when dividing by void type: for $\lambda_v<-1$, the improvement is 12, 20, 48, and 2~$\%$ for the noise-free, CMB-S4-like, SO, and {\it Planck}, respectively, and for $\lambda_v>7$, it is 4, 12, 43, and 13~$\%$. The trend shows that there is an improvement of the S/N as the noise level increases, except for {\it Planck}. We attribute this behaviour to the trade-off between reducing the intrinsic profile variance---by grouping similar void profiles---and the loss of statistical power from having fewer objects per bin. For S4-like and SO, grouping by void type outweighs the loss in statistics but for {\it Planck}, the gain is negligible in comparison with the loss in statistical power, resulting in a reduction of the S/N.\\

Dividing by type for the 2D voids shows no S/N improvement in either method compared to the full void sample. As we have already noted, in the non-rescaled approach, the reason could be the chosen void type binning. 2D bins were defined to have equal numbers of voids, which resulted in a wide range of radii (for example, the largest bin ranges from 32 to 283$h^{-1}~$Mpc). This broad range mixes very different profiles, which lowers the S/N for the non-rescaled method. The binning strategy works well for the rescaled method where stacking voids of different sizes does not significantly degrade the signal, but it is not optimal for a non-rescaled analysis.\\

\begin{table*}[ht!]
\centering
\begin{adjustbox}{width=\textwidth}
\begin{tabular}{lccccccccccccccc}
 & \multicolumn{1}{l}{} & \multicolumn{1}{l}{} & \multicolumn{1}{l}{} & \multicolumn{12}{c}{Noise configurations} \\ \cline{5-16} 
 & \multicolumn{3}{l|}{\multirow{4}{*}{}} & \multicolumn{1}{l}{} & \multicolumn{1}{l}{} & \multicolumn{1}{l|}{} & \multicolumn{1}{l}{} & \multicolumn{1}{l}{} & \multicolumn{1}{l|}{} & \multicolumn{1}{l}{} & \multicolumn{1}{l}{} & \multicolumn{1}{l|}{} & \multicolumn{1}{l}{} & \multicolumn{1}{l}{} & \multicolumn{1}{l|}{} \\
 & \multicolumn{3}{l|}{} & \multicolumn{3}{c|}{Noise-free} & \multicolumn{3}{c|}{S4-like} & \multicolumn{3}{c|}{SO} & \multicolumn{3}{c|}{Planck} \\
 & \multicolumn{3}{l|}{} & \multicolumn{1}{l}{} & \multicolumn{1}{l}{} & \multicolumn{1}{l|}{} & \multicolumn{1}{l}{} & \multicolumn{1}{l}{} & \multicolumn{1}{l|}{} & \multicolumn{1}{l}{} & \multicolumn{1}{l}{} & \multicolumn{1}{l|}{} & \multicolumn{1}{l}{} & \multicolumn{1}{l}{} & \multicolumn{1}{l|}{} \\ \cline{5-16} 
 & \multicolumn{3}{l|}{} & S/N & sm$_{\rm map}$ {[}°{]} & \multicolumn{1}{c|}{$\Delta(\rm S/N)_0$ {[}\%{]}} & S/N & sm$_{\rm map}$ {[}°{]} & \multicolumn{1}{c|}{$\Delta(\rm S/N)_0$ {[}\%{]}} & S/N & sm$_{\rm map}$ {[}°{]} & \multicolumn{1}{c|}{$\Delta(\rm S/N)_0$ {[}\%{]}} & S/N & sm$_{\rm map}$ {[}°{]} & \multicolumn{1}{c|}{$\Delta(\rm S/N)_0$ {[}\%{]}} \\ \cline{2-16} 
\multicolumn{1}{l|}{\multirow{33}{*}{\rotatebox{90}{Void type}}} & \multicolumn{1}{c|}{\multirow{11}{*}{3D Voids}} &  & \multicolumn{1}{c|}{} &  &  & \multicolumn{1}{c|}{} &  &  & \multicolumn{1}{c|}{} &  &  & \multicolumn{1}{c|}{} &  &  & \multicolumn{1}{c|}{} \\
\multicolumn{1}{l|}{} & \multicolumn{1}{c|}{} & \multirow{4}{*}{R} & \multicolumn{1}{c|}{all} & 20.58 & 0.2 & \multicolumn{1}{c|}{4.19} & 18.43 & 0.5 & \multicolumn{1}{c|}{13.15} & 12.95 & 0.5 & \multicolumn{1}{c|}{26.24} & 7.75 & 1 & \multicolumn{1}{c|}{26.97} \\
\multicolumn{1}{l|}{} & \multicolumn{1}{c|}{} &  & \multicolumn{1}{c|}{bin1} & 19.05 & 0.5 & \multicolumn{1}{c|}{8.76} & 18.14 & 0.5 & \multicolumn{1}{c|}{9.65} & 15.96 & 0.8 & \multicolumn{1}{c|}{17.16} & 11.58 & 1.2 & \multicolumn{1}{c|}{25.51} \\
\multicolumn{1}{l|}{} & \multicolumn{1}{c|}{} &  & \multicolumn{1}{c|}{bin2} & 12.73 & 0.2 & \multicolumn{1}{c|}{0.42} & 10.76 & 0.2 & \multicolumn{1}{c|}{7.06} & 7.62 & 0.2 & \multicolumn{1}{c|}{14.38} & 6.33 & 0.2 & \multicolumn{1}{c|}{15.89} \\
\multicolumn{1}{l|}{} & \multicolumn{1}{c|}{} &  & \multicolumn{1}{c|}{bin3} & 16.7 & 0.2 & \multicolumn{1}{c|}{4.45} & 15.75 & 0.2 & \multicolumn{1}{c|}{5.03} & 13.44 & 0.5 & \multicolumn{1}{c|}{2.93} & 9.81 & 0.2 & \multicolumn{1}{c|}{2.34} \\
\multicolumn{1}{l|}{} & \multicolumn{1}{c|}{} &  & \multicolumn{1}{c|}{} &  &  & \multicolumn{1}{c|}{} &  &  & \multicolumn{1}{c|}{} &  &  & \multicolumn{1}{c|}{} &  &  & \multicolumn{1}{c|}{} \\
\multicolumn{1}{l|}{} & \multicolumn{1}{c|}{} & \multirow{4}{*}{NR} & \multicolumn{1}{c|}{all} & 16.58 & 0.2 & \multicolumn{1}{c|}{2.99} & 14.66 & 0.2 & \multicolumn{1}{c|}{7.87} & 10.12 & 0.5 & \multicolumn{1}{c|}{11.73} & 10.33 & 1.2 & \multicolumn{1}{c|}{57.86} \\
\multicolumn{1}{l|}{} & \multicolumn{1}{c|}{} &  & \multicolumn{1}{c|}{bin1} & 18.58 & 0.2 & \multicolumn{1}{c|}{2.38} & 17.64 & 0.2 & \multicolumn{1}{c|}{1.93} & 14.96 & 0.2 & \multicolumn{1}{c|}{1.74} & 10.54 & 0.2 & \multicolumn{1}{c|}{2.53} \\
\multicolumn{1}{l|}{} & \multicolumn{1}{c|}{} &  & \multicolumn{1}{c|}{bin2} & 11.69 & 0 & \multicolumn{1}{c|}{0.0} & 10.34 & 0.2 & \multicolumn{1}{c|}{2.43} & 8.05 & 0.2 & \multicolumn{1}{c|}{1.11} & 7.51 & 0 & \multicolumn{1}{c|}{0.0} \\
\multicolumn{1}{l|}{} & \multicolumn{1}{c|}{} &  & \multicolumn{1}{c|}{bin3} & 17.24 & 0.2 & \multicolumn{1}{c|}{1.84} & 16.47 & 0.2 & \multicolumn{1}{c|}{3.25} & 14.48 & 0.5 & \multicolumn{1}{c|}{3.29} & 11.7 & 1.2 & \multicolumn{1}{c|}{14.54} \\
\multicolumn{1}{l|}{} & \multicolumn{1}{c|}{} &  & \multicolumn{1}{c|}{} &  &  & \multicolumn{1}{c|}{} &  &  & \multicolumn{1}{c|}{} &  &  & \multicolumn{1}{c|}{} &  &  & \multicolumn{1}{c|}{} \\ \cline{2-16} 
\multicolumn{1}{l|}{} & \multicolumn{1}{c|}{\multirow{11}{*}{\begin{tabular}[c]{@{}c@{}}2D Voids\\ \tiny{(sm$_{\rm VF}$ = 10 $h^{-1}$ Mpc)}\end{tabular}}} &  & \multicolumn{1}{c|}{} &  &  & \multicolumn{1}{c|}{} &  &  & \multicolumn{1}{c|}{} &  &  & \multicolumn{1}{c|}{} &  &  & \multicolumn{1}{c|}{} \\
\multicolumn{1}{l|}{} & \multicolumn{1}{c|}{} & \multirow{4}{*}{R} & \multicolumn{1}{c|}{all} & 25.01 & 0.2 & \multicolumn{1}{c|}{2.12} & 23.39 & 0.2 & \multicolumn{1}{c|}{3.77} & 19.22 & 0.5 & \multicolumn{1}{c|}{14.4} & 12.23 & 0.8 & \multicolumn{1}{c|}{38.78} \\
\multicolumn{1}{l|}{} & \multicolumn{1}{c|}{} &  & \multicolumn{1}{c|}{bin1} & 16.78 & 0.2 & \multicolumn{1}{c|}{1.06} & 15.52 & 0.2 & \multicolumn{1}{c|}{1.48} & 11.5 & 0.2 & \multicolumn{1}{c|}{5.25} & 7.86 & 0.8 & \multicolumn{1}{c|}{31.73} \\
\multicolumn{1}{l|}{} & \multicolumn{1}{c|}{} &  & \multicolumn{1}{c|}{bin2} & 15.87 & 0 & \multicolumn{1}{c|}{0.0} & 14.63 & 0.2 & \multicolumn{1}{c|}{1.12} & 11.99 & 0.5 & \multicolumn{1}{c|}{2.95} & 8.51 & 0.5 & \multicolumn{1}{c|}{5.67} \\
\multicolumn{1}{l|}{} & \multicolumn{1}{c|}{} &  & \multicolumn{1}{c|}{bin3} & 18.2 & 0.2 & \multicolumn{1}{c|}{1.51} & 17.66 & 0.2 & \multicolumn{1}{c|}{3.24} & 15.5 & 0.5 & \multicolumn{1}{c|}{9.24} & 11.01 & 0.5 & \multicolumn{1}{c|}{14.81} \\
\multicolumn{1}{l|}{} & \multicolumn{1}{c|}{} &  & \multicolumn{1}{c|}{} &  &  & \multicolumn{1}{c|}{} &  &  & \multicolumn{1}{c|}{} &  &  & \multicolumn{1}{c|}{} &  &  & \multicolumn{1}{c|}{} \\
\multicolumn{1}{l|}{} & \multicolumn{1}{c|}{} & \multirow{4}{*}{NR} & \multicolumn{1}{c|}{all} & 22.17 & 0 & \multicolumn{1}{c|}{0.0} & 21.03 & 0.2 & \multicolumn{1}{c|}{0.32} & 16.67 & 0.2 & \multicolumn{1}{c|}{0.88} & 10.54 & 0.2 & \multicolumn{1}{c|}{1.03} \\
\multicolumn{1}{l|}{} & \multicolumn{1}{c|}{} &  & \multicolumn{1}{c|}{bin1} & 16.35 & 0 & \multicolumn{1}{c|}{0.0} & 15.37 & 0 & \multicolumn{1}{c|}{0.0} & 11.6 & 0 & \multicolumn{1}{c|}{0.0} & 7.41 & 0 & \multicolumn{1}{c|}{0.0} \\
\multicolumn{1}{l|}{} & \multicolumn{1}{c|}{} &  & \multicolumn{1}{c|}{bin2} & 15.82 & 0.2 & \multicolumn{1}{c|}{0.16} & 14.54 & 0.2 & \multicolumn{1}{c|}{1.21} & 11.43 & 0.2 & \multicolumn{1}{c|}{0.67} & 7.67 & 0 & \multicolumn{1}{c|}{0.0} \\
\multicolumn{1}{l|}{} & \multicolumn{1}{c|}{} &  & \multicolumn{1}{c|}{bin3} & 16.0 & 0.2 & \multicolumn{1}{c|}{0.26} & 15.7 & 0.2 & \multicolumn{1}{c|}{1.59} & 13.83 & 0.5 & \multicolumn{1}{c|}{3.19} & 10.18 & 0.5 & \multicolumn{1}{c|}{4.73} \\
\multicolumn{1}{l|}{} & \multicolumn{1}{c|}{} &  & \multicolumn{1}{c|}{} &  &  & \multicolumn{1}{c|}{} &  &  & \multicolumn{1}{c|}{} &  &  & \multicolumn{1}{c|}{} &  &  & \multicolumn{1}{c|}{} \\ \cline{2-16} 
\multicolumn{1}{l|}{} & \multicolumn{1}{c|}{\multirow{11}{*}{\begin{tabular}[c]{@{}c@{}}2D Voids\\ \tiny{(sm$_{\rm VF}$ = 5 $h^{-1}$ Mpc)}\end{tabular}}} &  & \multicolumn{1}{c|}{} &  &  & \multicolumn{1}{c|}{} &  &  & \multicolumn{1}{c|}{} &  &  & \multicolumn{1}{c|}{} &  &  & \multicolumn{1}{c|}{} \\
\multicolumn{1}{l|}{} & \multicolumn{1}{c|}{} & \multirow{4}{*}{R} & \multicolumn{1}{c|}{all} & 34.38 & 0 & \multicolumn{1}{c|}{0.0} & 30.52 & 0.2 & \multicolumn{1}{c|}{3.14} & 21.63 & 0.5 & \multicolumn{1}{c|}{7.56} & 12.59 & 0.8 & \multicolumn{1}{c|}{26.24} \\
\multicolumn{1}{l|}{} & \multicolumn{1}{c|}{} &  & \multicolumn{1}{c|}{bin1} & 24.31 & 0 & \multicolumn{1}{c|}{0.0} & 21.33 & 0.2 & \multicolumn{1}{c|}{1.9} & 13.51 & 0.5 & \multicolumn{1}{c|}{4.48} & 7.22 & 0 & \multicolumn{1}{c|}{0.0} \\
\multicolumn{1}{l|}{} & \multicolumn{1}{c|}{} &  & \multicolumn{1}{c|}{bin2} & 20.75 & 0.2 & \multicolumn{1}{c|}{1.06} & 18.92 & 0.2 & \multicolumn{1}{c|}{0.45} & 13.88 & 0.5 & \multicolumn{1}{c|}{1.43} & 8.44 & 0.5 & \multicolumn{1}{c|}{4.71} \\
\multicolumn{1}{l|}{} & \multicolumn{1}{c|}{} &  & \multicolumn{1}{c|}{bin3} & 23.46 & 0.2 & \multicolumn{1}{c|}{1.37} & 22.22 & 0.2 & \multicolumn{1}{c|}{4.24} & 17.32 & 0.2 & \multicolumn{1}{c|}{4.89} & 11.49 & 0.8 & \multicolumn{1}{c|}{8.4} \\
\multicolumn{1}{l|}{} & \multicolumn{1}{c|}{} &  & \multicolumn{1}{c|}{} &  &  & \multicolumn{1}{c|}{} &  &  & \multicolumn{1}{c|}{} &  &  & \multicolumn{1}{c|}{} &  &  & \multicolumn{1}{c|}{} \\
\multicolumn{1}{l|}{} & \multicolumn{1}{c|}{} & \multirow{4}{*}{NR} & \multicolumn{1}{c|}{all} & 28.33 & 0.2 & \multicolumn{1}{c|}{0.67} & 25.72 & 0.2 & \multicolumn{1}{c|}{2.91} & 18.05 & 0.2 & \multicolumn{1}{c|}{4.74} & 10.75 & 1 & \multicolumn{1}{c|}{17.47} \\
\multicolumn{1}{l|}{} & \multicolumn{1}{c|}{} &  & \multicolumn{1}{c|}{bin1} & 23.24 & 0.2 & \multicolumn{1}{c|}{1.51} & 21.06 & 0.2 & \multicolumn{1}{c|}{3.56} & 13.69 & 0.2 & \multicolumn{1}{c|}{5.82} & 8.12 & 0.5 & \multicolumn{1}{c|}{16.86} \\
\multicolumn{1}{l|}{} & \multicolumn{1}{c|}{} &  & \multicolumn{1}{c|}{bin2} & 20.12 & 0.2 & \multicolumn{1}{c|}{0.49} & 18.42 & 0.2 & \multicolumn{1}{c|}{1.88} & 13.19 & 0.2 & \multicolumn{1}{c|}{3.88} & 8.12 & 1.2 & \multicolumn{1}{c|}{21.53} \\
\multicolumn{1}{l|}{} & \multicolumn{1}{c|}{} &  & \multicolumn{1}{c|}{bin3} & 20.41 & 0.2 & \multicolumn{1}{c|}{2.61} & 19.13 & 0.2 & \multicolumn{1}{c|}{3.83} & 15.18 & 0.2 & \multicolumn{1}{c|}{2.93} & 10.94 & 1.2 & \multicolumn{1}{c|}{7.11} \\
\multicolumn{1}{l|}{} & \multicolumn{1}{c|}{} &  & \multicolumn{1}{c|}{} &  &  & \multicolumn{1}{c|}{} &  &  & \multicolumn{1}{c|}{} &  &  & \multicolumn{1}{c|}{} &  &  & \multicolumn{1}{c|}{} \\ \cline{2-16} 
 & \multicolumn{1}{l}{} & \multicolumn{1}{l}{} & \multicolumn{1}{l}{} & \multicolumn{1}{l}{} & \multicolumn{1}{l}{} & \multicolumn{1}{l}{} & \multicolumn{1}{l}{} & \multicolumn{1}{l}{} & \multicolumn{1}{l}{} & \multicolumn{1}{l}{} & \multicolumn{1}{l}{} & \multicolumn{1}{l}{} & \multicolumn{1}{l}{} & \multicolumn{1}{l}{} & \multicolumn{1}{l}{}
\end{tabular}
\end{adjustbox}
\caption{Signal-to-noise ratios for 3D and 2D voids under different rescaling assumptions (R = Rescaled, NR = Not rescaled) and experimental noise configurations. Smoothing scales are given in degrees. The column $\Delta( S/N)_0$ shows the relative change of S/N compared to the baseline value at zero smoothing. The different bins correspond to $\lambda$ $\in$ [-70,70] (all), [-70,-1], [1,7], and [7,70] for 3D voids, $R_{\rm v}$ $\in$ [11,283] (all), [11,28], [28,49], and [49,283]~$h^{-1}$~Mpc for 2D voids with ${\rm sm_{VF}}=10~h^{-1}$~Mpc, and $R_{\rm v}$ $\in$ [5,283] (all), [5,16], [16,32], and [32,283]~$h^{-1}$~Mpc for 2D voids with ${\rm sm_{VF}}=5~h^{-1}$~Mpc.}
\label{tab:comparison_SNR}
\end{table*}

\begin{figure*}[!b]
    \centering
    \includegraphics[width=15cm]{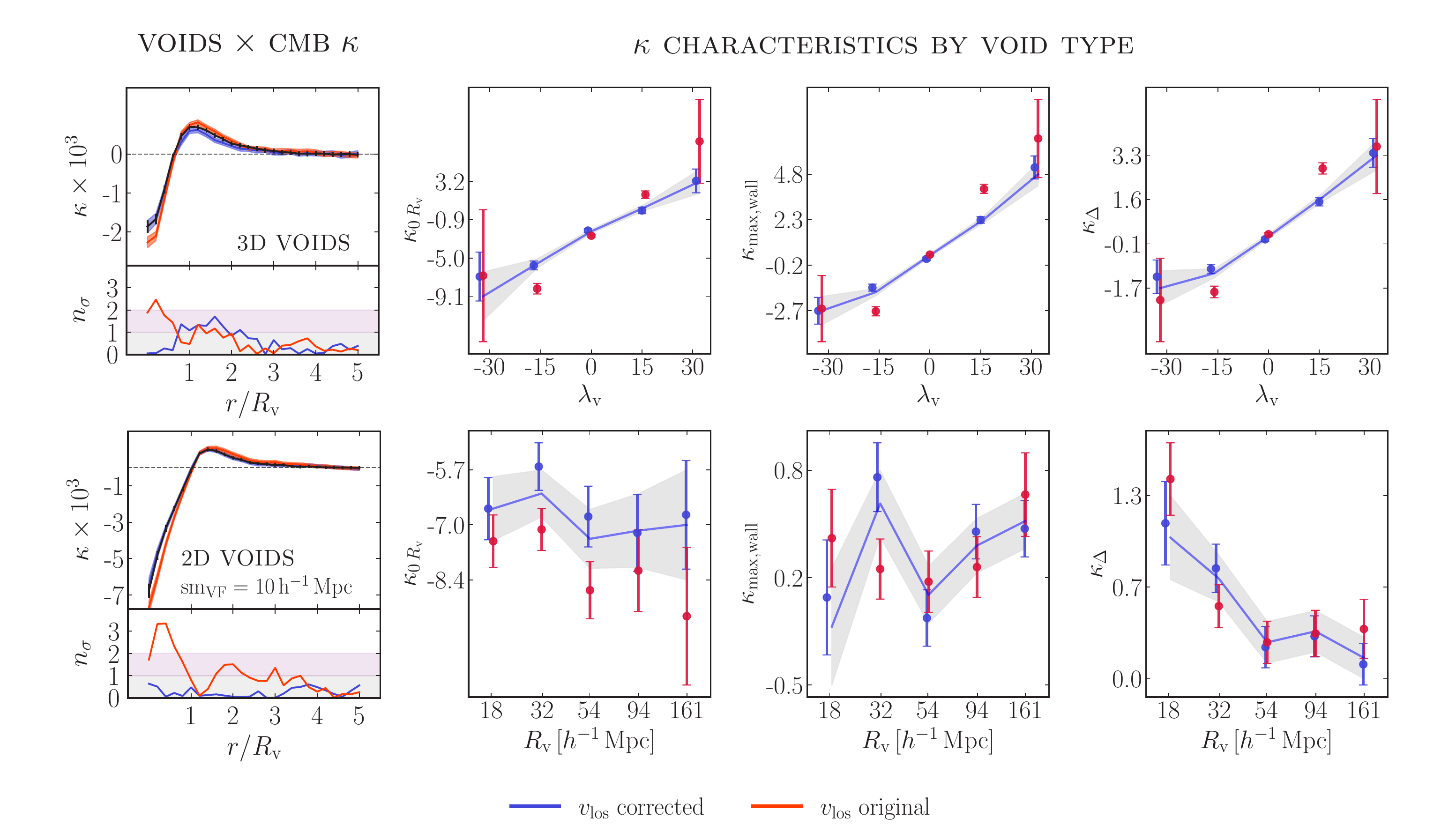} 
    \caption{Void $\times$ CMB lensing profiles and profile parameters ($\rm \kappa_{0R_{\rm v}}$, $\rm \kappa_{max,wall}$, $\rm \kappa_\Delta$) for \textsc{Agora} with original vs. sign-flipped velocities. Residuals are relative to the mean \emph{Roman}-consistent profiles. In this plot we are just considering a noise-free CMB $\kappa$ map. 
}
    \label{fig:testzH}
\end{figure*}
\section{AGORA peculiar velocities}
\label{sec:app_agora_velocities}
In the first of these series of papers, \citep{perezsar2025_roman_mocks}, we reported that the line-of-sight velocities in the \textsc{Agora} simulation suffered from a sign inversion. When applying the \emph{analog matching} method to the affected halo catalogs, the results still reproduce the galaxy power spectrum of the reference \emph{Roman} mock, but they deliver unphysical void statistics (see Figure D.1 in the Appendix of \citealt{perezsar2025_roman_mocks}), which highlight the importance of using void statistics as a diagnostic for simulation consistency. 

Here we test how this inversion impacts the lensing signal. The flipped-velocity catalogs create an artificially deep convergence profile, deviating by up to $3\sigma$ from the expected amplitude. The trend follows void type: voids-in-voids show a stronger, spurious signal, while voids-in-clouds appear shallower. 

We stress that while Void $\times$ CMB lensing is robust against moderate variations in galaxy bias, it is sensitive to the selection effects that define the void sample. As illustrated in this example, a mock can successfully reproduce the clustering statistics while delivering unphysical void properties and lensing profiles. In some forward-modelling approaches, such as Halo Occupation Distribution models that force-match observed clustering, the tracer population may become partially decoupled from the underlying matter distribution. By absorbing physical effects into the tracer bias to reproduce the observations, these approaches can produce void catalogs that fail to capture the true depth of matter underdensities, ultimately leading to biased lensing amplitudes and obscured cosmological signatures.  Thus, we consider that to ensure robust cosmological constraints, void statistics must be used as a diagnostic tool to verify that the mock captures both the true depth of the matter underdensities and the spatial distribution of the tracers.

\end{document}